\documentclass{aastex61}

\newcommand\kms {$\,$km$\,$s$^{-1}$}
\newcommand\lsols {$\,$L$_{\odot}$}

\newcommand\mjyb {$\,$mJy$\,$beam$^{-1}$}

\received{September 4, 2017}
\revised{September 4, 2017}
\accepted{September 7, 2017}
\submitjournal{ApJS}
\shorttitle{Long-Term Variability of H$_2$CO Masers}
\shortauthors{Andreev et al.}
\begin{document}

\title{Long-Term Variability of H$_2$CO Masers in Star-Forming Regions}

\correspondingauthor{E. D. Araya}
\email{ed-araya@wiu.edu}

\author[0000-0002-7098-6480]{N. Andreev}

\author{E. D. Araya}
\affiliation{Physics Department, Western Illinois University, 1 University Circle, Macomb, IL 61455, USA.}

\author{I.\ M.\ Hoffman}
\affiliation{Quest University, 3200 University Boulevard, Squamish, British Columbia V8B0N8, Canada.}

\author{P. Hofner}
\altaffiliation{Adjunct Astronomer at the National Radio Astronomy Observatory, 1003 Lopezville Road, Socorro, NM 87801, USA.}
\affiliation{New Mexico Institute of Mining and Technology, Physics Department, 801 Leroy Place, Socorro, NM 87801, USA.}

\author{S. Kurtz}
\affiliation{Instituto de Radioastronom\'{\i}a y Astrof\'{\i}sica, Universidad Nacional Aut\'onoma de M\'exico,  Apdo. Postal 3-72, 58090, Morelia, Michoac\'an, Mexico.}

\author{H. Linz}
\affiliation{Max--Planck--Institut f\"ur Astronomie, K\"onigstuhl 17, D--69117, Heidelberg, Germany.}

\author{L. Olmi}
\affiliation{INAF, Osservatorio Astrofisico di Arcetri, Largo E. Fermi 5, I-50125, Firenze, Italy.}
\affiliation{University of Puerto Rico, R\'{\i}o Piedras Campus, Physics Dept., Box 23343, UPR Station, San Juan, PR 00931, USA.}

\author{I. Lorran-Costa}
\affiliation{Physics Department, Western Illinois University, 1 University Circle, Macomb, IL 61455, USA.}

\begin{abstract}
We present results of a multi-epoch monitoring program on variability of 6$\,$cm formaldehyde (H$_2$CO) masers in the massive star forming region NGC$\,$7538$\,$IRS$\,$1 from 2008 to 2015 conducted with the GBT, WSRT, and VLA. We found that the similar variability behaviors of the two formaldehyde maser velocity components in NGC$\,$7538$\,$IRS$\,$1 (which was pointed out by Araya and collaborators in 2007) have continued. The possibility that the variability is caused by changes in the maser amplification path in regions with similar morphology and kinematics is discussed. We also observed 12.2$\,$GHz methanol and 22.2$\,$GHz water masers toward NGC$\,$7538$\,$IRS$\,$1. The brightest maser components of CH$_3$OH and H$_2$O species show a decrease in flux density as a function of time. The brightest H$_2$CO maser component also shows a decrease in flux density and has a similar LSR velocity to the brightest H$_2$O and 12.2$\,$GHz CH$_3$OH masers. The line parameters of radio recombination lines and the 20.17 and 20.97$\,$GHz CH$_3$OH transitions in NGC$\,$7538$\,$IRS$\,$1 are also reported. In addition, we observed five other 6$\,$cm formaldehyde maser regions. We found no evidence of significant variability of the 6$\,$cm masers in these regions with respect to previous observations, the only possible exception being the maser in G29.96$-$0.02. All six sources were also observed in the H$_2^{13}$CO isotopologue transition of the 6$\,$cm H$_2$CO line; H$_2^{13}$CO absorption was detected in five of the sources. Estimated column density ratios [H$_2^{12}$CO]/[H$_2^{13}$CO] are reported.

\end{abstract}


\section{Introduction} \label{sec:intro}

Since their discovery in 1965 \citep{1965Natur.208...29W}, astrophysical masers have been the object of active research. They have been detected in different regions, including clouds of water vapor around moons of Saturn \citep{2009A&A...494L...1P}, atmospheres of comets  \citep{1974A&A....34..163B,1974ApJ...189L.137T,1994A&A...287..647B}, circumstellar envelopes of evolved stars (e.g.,  \citealt{1968Sci...161..778W}), supernova remnants (e.g.,   \citealt{1994ApJ...424L.111F}), and star forming regions (e.g., \citealt{1967ARA&A...5..183R}). In addition, masers and megamasers have been found in extragalactic environments (e.g.,  \citealt{1985Natur.315...26B};  \citealt{1994ApJ...437L..35H};  \citealt{2005ARA&A..43..625L};  \citealt{2010ApJ...724L.158S}).
There are several types of masers which have been detected exclusively in high-mass star forming regions (HMSFR) including  6.7 and 12.2$\,$GHz methanol and 6$\,$cm (4.83$\,$GHz) formaldehyde masers (e.g., \citealt{1998MNRAS.301..640W}; \citealt{2013MNRAS.435..524B}; \citealt{2007IAUS..242..110A}; \citealt{2015ApJS..221...10A}). These masers are not only signposts of high-mass star formation but also help trace small-scale dynamics, morphology, and the locations of outflows and accretion disks (e.g., \citealt{2009A&A...502..155B}; \citealt{2010A&A...517A..78S}; \citealt{2011A&A...535L...8G}; \citealt{2011MNRAS.410..627T}; \citealt{2011A&A...536A..38M}).

Astrophysical masers exhibit diverse temporal variability with time scales ranging from less than a day to years, from aperiodic flares to quasi-periodic and sinusoidal patterns (e.g., \citealt{2007A&A...476..373F}; \citealt{2004MNRAS.355..553G}; \citealt{2012IAUS..287...85G}; \citealt{2014MNRAS.437.1808G}). Excluding stellar masers (i.e., associated with evolved stars), maser variability is not yet well understood. In some cases, flux density variability of certain maser species  (6.7 and 12.2$\,$GHz methanol, 6$\,$cm formaldehyde, 6.035 GHz hydroxyl, and 22.2$\,$GHz water masers) in HMSFR is predictable, such as monotonic increase/decrease in flux density and periodic maser flares (e.g., \citealt{2003MNRAS.339L..33G}; \citealt{2010ApJ...717L.133A}; \citealt{2012ApJ...750..170A}; \citealt{2016MNRAS.456.4335M}). A particularly interesting recent example is the anti-correlated flares of methanol and water masers in the intermediate-mass young stellar object G107.298+5.639, which could be tracing periodic accretion instabilities in a protobinary disk (\citealt{2016MNRAS.459L..56S}). The study of correlated variability of different maser species and velocity components of a single maser transition is aimed to investigate the mechanism of maser excitation and to use masers as astrophysical probes to better understand the processes of intermediate and high-mass star formation.

The 6$\,$cm line of formaldehyde (H$_2$CO) is a rare maser transition that has been reported toward nine HMSFRs in our Galaxy: NGC$\,$7538$\,$IRS$\,$1 \citep{1974ApJ...191L..77D, 2003ApJ...598.1061H}, Sgr B2 \citep{1983MNRAS.205P..27W, 1994ApJ...434..237M, 2007ApJ...654..971H}, G29.96$-$0.02 \citep{1994ApJ...430L.129P, 2003ApJ...598.1061H}, IRAS$\,$18566+0408 (G37.55+0.20; \citealt{2004ApJS..154..579A}), G23.71$-$0.20 \citep{2006ApJ...643L..33A}, G23.01$-$0.41 \citep{2008ApJS..178..330A}, G25.83$-$0.18 \citep{2008ApJS..178..330A}, G32.74$-$0.07 \citep{2015ApJS..221...10A}, and G0.38+0.04 \citep{2015A&A...573A.106G}, and three extragalactic regions (\citealt{1986ApJ...305..830B}; \citealt{2004ApJS..154..541A}; see however \citealt{2008ApJ...673..832M}). 

\begin{figure}
\includegraphics{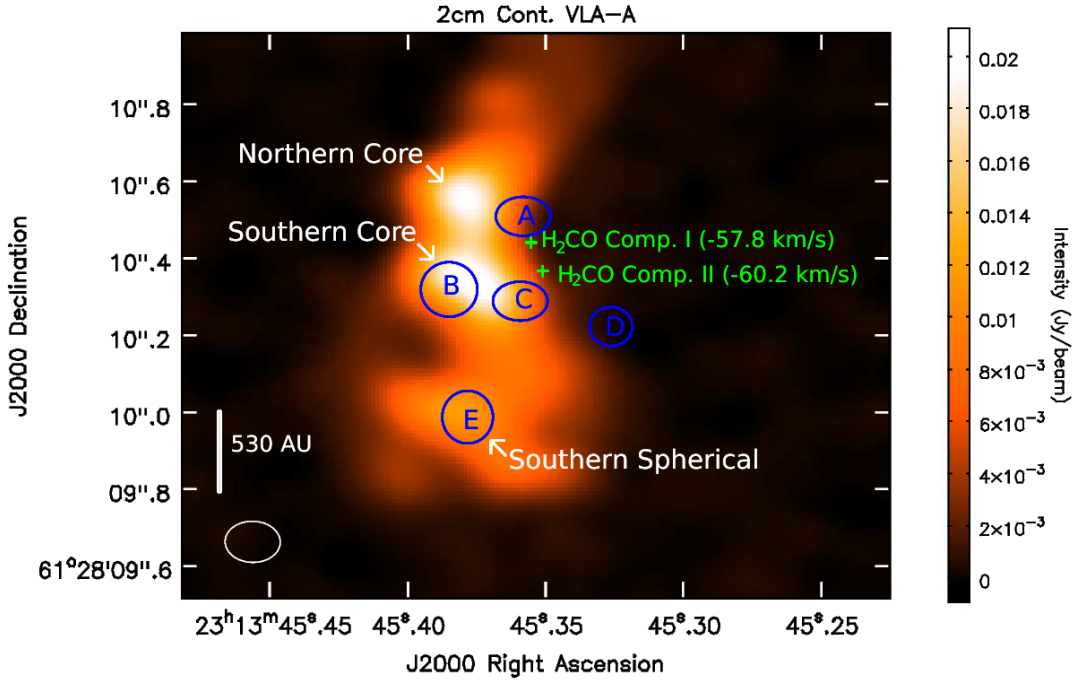}
\vspace*{11cm}
\caption{NGC$\,$7538$\,$IRS$\,$1 region; 2$\,$cm radio continuum is shown in colors (VLA A-configuration, 2$\,$cm Ku-band observations, VLA archive, project AM0169). The crosses show the location of the 6$\,$cm H$_2$CO masers \citep{2003ApJ...598.1061H}. The blue ellipses show the approximate location of 6.7$\,$GHz CH$_3$OH maser groups. The identification of the CH$_3$OH maser groups (A to E) and the radio continuum sources (Northern Core, Southern Core, and Southern Spherical) are as presented by \citet{2014A&A...566A.150M}.  \label{fig:ngc7538_cont}}
\end{figure}

The first 6$\,$cm formaldehyde maser was detected in NGC$\,$7538$\,$IRS$\,$1 \citep{1974ApJ...191L..77D, 1981ApJ...245L..15R}. This region is located at a distance of 2.65$^{+0.12}_{-0.11}\,$kpc \citep{2009ApJ...693..406M} and it is one of the richest maser sites known (e.g., \citealt{2010ApJ...713..423G}; \citealt{2011A&A...533A..47S}; \citealt{2011AJ....142..202H}). Interferometric observations of the double-peaked 6$\,$cm formaldehyde maser in NGC$\,$7538$\,$IRS$\,$1 show that the two velocity components originate from two regions oriented approximately NE-SW (Figure~\ref{fig:ngc7538_cont}), which are separated by $\sim$200$\,$AU in the plane of the sky (based on \citealt{2003ApJ...598.1061H} and the distance reported by \citealt{2009ApJ...693..406M}). VLBA observations by \citet{2003ApJ...598.1061H} show that the NE maser (Component I) has a velocity gradient of $1$\kms~over a projected scale of $\sim 100\,$AU. The formaldehyde masers could be located inside of (or close to) a possible circumbinary envelope that has been proposed by \citet{2015A&A...573A.108G}. While the H$_2$CO maser Component II seems to reside approximately between two 6.7 GHz methanol maser clusters (A and C, Figure~\ref{fig:ngc7538_cont}; see also \citealt{2014A&A...566A.150M}), Component I is very close (in projection) to the northern 6.7 GHz maser cluster A, which shows small-scale velocity gradients in thermal molecular lines from NH$_3$ and CH$_3$OH \citep{2017arXiv170506246B}.

The two formaldehyde masers did not show significant variability over a period of about eight years after their discovery. After 1982, the intensity of Component I increased while Component II showed no variability; approximately 14 years later, the intensity of Component II started to increase. After the onset of variability, both components showed similar variability slopes, which suggested a connection between the flux density changes of the maser regions \citep{2007ApJS..170..152A}. 

In this paper we report monitoring observations of the 6$\,$cm formaldehyde masers in NGC$\,$7538 IRS$\,$1 to investigate whether the similar variability trend of both components has continued. We also investigate variability of 6$\,$cm formaldehyde masers in five other star forming regions. We report detection of absorption of 4.59$\,$GHz H$_2^{13}$CO in five of the six HMSFRs in our sample, as well as observations of 12.2$\,$GHz methanol masers, 22.2$\,$GHz water masers, and 20.17 and 20.97$\,$GHz methanol absorption lines in NGC$\,$7538$\,$IRS$\,$1.\footnote{This work is based on Chapter 3 of the MS Thesis ``From Diffuse Clouds to Massive Star Forming Regions'' by N. Andreev, WIU, 2014.}

\section{Observations}

\subsection{Robert C. Byrd Green Bank Telescope Observations}

We used the Robert C. Byrd Green Bank Telescope (GBT, Green Bank Observatory) in West Virginia in our monitoring program to study the long term variability of formaldehyde masers in NGC$\,$7538$\,$IRS$\,$1. The pointing position of all observations was R.A. = $23^h$13$^m$45.36$^s$, Decl.= $+61^o28'10.5$\arcsec (J2000; \citealt{2010ApJ...713..423G}). Doppler tracking was centered at a LSR velocity of $-$55.00~\kms. All data reduction was done using GBTIDL routines provided by NRAO. We report GBT observations from May 2008 to May 2011 of NGC$\,$7538$\,$IRS$\,$1 in the C (6$\,$cm), Ku (2$\,$cm), and K (1.3$\,$cm) bands. We used the GBT spectrometer with 4 spectral windows (see Table \ref{tab:tbl-1}). After calibration, orthogonal polarization spectra were checked for consistency and averaged. A linear baseline was subtracted. Some spectra were smoothed to improve signal-to-noise; the final channel separations after smoothing are listed in Table \ref{tab:tbl-1}.

\begin{deluxetable}{lclccc}
\tablecaption{Spectral Lines \label{tab:tbl-1}}
\tablewidth{0pt}
\tablehead{
\colhead{Molecule} & \colhead{Transition$^*$} & \colhead{Rest Frequency$^*$} &
\colhead{Epoch/Mode$^+$} & \colhead{Chan. Spacing} & \colhead{RRLs in Bandpass} \\
\colhead{ } & \colhead{ } & \colhead{(GHz)} & 
\colhead{ } & \colhead{(\kms)} & \colhead{ }
}
\startdata
H$_2$C$^{18}$O  & 1(1,0)-1(1,1)                  & 4.3887970  & 1a, 2c, 3c, 4ac, 5c, 6c  & 0.42            &   \\
H$_2^{13}$CO    & 1(1,0)-1(1,1)                  & 4.5930885  & 1a, 2c, 3c, 4ac, 5c, 6c  & 0.40$^1$        & H141$\beta$ \\
H$_2$CO	       & 1(1,0)-1(1,1)                  & 4.8296594  & 1a, 2c, 3c, 4ac, 5c, 6c  & 0.095/0.57$^1$  & H174$\delta^2$ \\
               &				&            & 7w, 8w                   & 0.095           &		\\
	       &				&	     & 9v                       & 0.061      	  &			\\
CH$_3$OH       & 3(1,2)-3(1,3) A$^{-+}$          & 5.0053208  & 1a, 2c, 3c, 4ac, 5c, 6c  & 0.37            & H137$\beta$, H109 \\
               &                                &            &                          &                 &   \\
CH$_3$OH       & 2(0,2)-3(-1,3) E               & 12.1785970 & 1b, 5b                   & 0.075           &   \\
H$_2^{13}$CO    & 2(1,1)-2(1,2)                  & 13.7788041 & 1b, 5b                   & 0.27            &   \\
H$_2$CO        & 2(1,1)-2(1,2)                  & 14.4884801 & 1b, 5b                   & 0.32            &   \\
$^{13}$CH$_3$OH & 2(0,2)-3(-1,3) E               & 14.7822120 & 1b, 5b                   & 0.25            &   \\
               &                                &            &                          &                 &   \\
CH$_3$OH       & 2(1,1)-3(0,3) E                & 19.9673961 & 1b, 2b, 5b               & 0.18            & H86$\beta^3$ \\
CH$_3$OH       & 11(1,11)-10(2,8) A$^{++}$           & 20.1710890 & 1b, 2b, 5b           & 0.18            &   \\
CH$_3$OH       & 10(1,10)-11(2,9) A$^{++}$ $\nu_t$=1 & 20.9706510 & 1b, 2b, 5b           & 0.35            &  \\
H$_2$O         & 6(1,6)-5(2,3)                  & 22.2350800 & 1b, 2b, 5b               & 0.16            &   \\
\enddata 
\tablenotetext{*}{Quantum numbers and rest frequencies obtained from
the Cologne Database for Molecular Spectroscopy \citep{2005JMoSt.742..215M},
the Lovas database (http://www.nist.gov/pml/data/micro/index.cfm), 
and the JPL Molecular Spectroscopy catalog \citep{1998JQSRT..60..883P} accessed through {\it {splatalogue}} (http://www.cv.nrao.edu/php/splat/).}
\tablenotetext{+}{~Observing epochs: 1. 2008-May-05; 2. 2009-May-18;
  3. 2009-July-03; 4. 2009-July-17; 5. 2010-May-07; 6. 2011-May-31; 7. 2014-December-07, 8. 2015-May-29, 9. 2015-June-28.
Observing Modes: a. GBT observations, frequency switching; b. GBT observations, beam switching; c. GBT observations, position switching; w. WSRT observations, v. VLA observations.}
\tablenotetext{1}{~Channel width of RRLs. The isotopologue spectra were smoothed to different channel widths: 0.8\kms~(G23.01$-$0.41, G23.71$-$0.20, IRAS$\,$18566+0408), 1.2\kms~(NGC$\,$7538$\,$IRS$\,$1), and 1.6\kms~(G25.83$-$0.18, G29.96$-$0.02).}
\tablenotetext{2}{~The H174$\delta$ was only observed in `c' mode (GBT observations, position switching).}
\tablenotetext{3}{~The H86$\beta$ line was only observed in Epoch 1 because a bandwidth of 50$\,$MHz was used in the K-Band observations. The line was outside the bandpass in the other epochs (2 and 5) because of a smaller bandwidth (12.5$\,$MHz).}
\end{deluxetable}

\begin{deluxetable}{lccccc}
\tablecaption{Observed Sources \label{tab:tbl-2}}
\tablewidth{0pt}
\tablehead{
\colhead{Name} & \colhead{$RA\, (J2000)$} & \colhead{$Decl.\, (J2000)$} &
\colhead{Epoch/Mode$^+$} & \colhead{Adopted Distance} & \colhead{Ref. Distance}  \\
\colhead{ } & \colhead{ (h m s)} & \colhead{($\degr ~ \arcmin ~ \arcsec$)} & \colhead{ } & \colhead{(kpc) }& \colhead{ }
}
\startdata
G23.01$-$0.41	         & 18 34 40.3  & $-$09 00 38   & 4a                   & 4.59$^{+0.38}_{-0.33}$     & 1 \\
G23.71$-$0.20	         & 18 35 12.4  & $-$08 17 39   & 4a                   & 6.21$^{+1.0}_{-0.80}$      & 2 \\
G25.83$-$0.18	         & 18 39 03.6  & $-$06 24 11   & 4a                   & 5.0$^{+0.3}_{-0.3}$         & 3 \\
G29.96$-$0.02	         & 18 46 03.8  & $-$02 39 22   & 4a                   & 5.3$^{+0.5}_{-0.5}$          & 4 \\
IRAS$\,$18566+0408       & 18 59 10.0  &  +04 12 15    & 1a and 4a            & 6.7$^{+1.4}_{-1.4}$        & 5 \\
NGC$\,$7538$\,$IRS$\,$1  & 23 13 45.36 &  +61 28 10.5  & 1ab, 2bc, 3c, 4c,    & 2.65$^{+0.12}_{-0.11}$     &  6 \\
                         &	       &               & 5bc, 6c, 7w, 8w, 9v  &	                       &   \\
\enddata 
\tablenotetext{+}{~Same notation as in Table~\ref{tab:tbl-1}.}
\tablecomments{ Reference of adopted distance:  
(1) \citet{2009ApJ...693..424B};
(2) \citet{2014ApJ...781..108S};
(3) \citet{2011MNRAS.417.2500G};
(4) \citet{2014ApJ...783..130R};
(5) \citet{2004ApJS..154..579A};
(6) \citet{2009ApJ...693..406M}.}
\end{deluxetable}

\subsubsection{C-Band Observations}

The GBT half-power beam width (HPBW) within the frequency range of our C-Band observations (4.3 to 5.0$\,$GHz) is between 2.9\arcmin~and 2.5\arcmin. We conducted six epochs of C-Band observations  (see Table \ref{tab:tbl-1}) of NGC$\,$7538$\,$IRS$\,$1: one in frequency switching mode and five in position switching mode. In addition, two observations of IRAS$\,$18566+0408 and a single observation of G23.01$-$0.41, G23.71$-$0.20, G25.83$-$0.18, and G29.96$-$0.02 were conducted (Table \ref{tab:tbl-2}). Specifically:

\renewcommand{\labelitemi}{$-$}
\begin{itemize} 

\item Epoch 1: The C-Band observations of NGC$\,$7538 IRS$\,$1 and IRAS$\,$18566+0408 were done in frequency switching mode ($\pm 3.125\,$MHz switching cycle), 12.5$\,$MHz ($\sim$ 300~\kms) bandwidth, 9 level sampling, 8192 channels (1.526$\,$kHz, 0.095~\kms~channel width), dual linear polarization.

\item Epochs 2 - 6: The C-Band observations of NGC$\,$7538 IRS$\,$1 were done in position switching mode to improve the baseline quality with respect to frequency switching observations. We used 12.5$\,$MHz bandwidth, 4096 channels (3.052~kHz, $\sim$ 0.19~\kms~initial channel separation), and dual linear polarization. 

\item Epoch 4: The C-Band observations of G23.01$-$0.41, G23.71$-$0.20, G25.83$-$0.18, G29.96$-$0.02, and IRAS$\,$18566 +0408 were done in frequency switching mode ($\pm 3.125\,$MHz switching cycle), 12.5$\,$MHz bandwidth, 9 level sampling, 8192 channels (1.526$\,$kHz), dual linear polarization.

\end{itemize}

For NGC$\,$7538$\,$IRS$\,$1 the total integration times on-source were 5 minutes for epochs 1 and 5; 10 minutes for epochs 2 and 4; 18 minutes for epoch 3; and approximately 3 minutes for epoch 6. For G25.83$-$0.18 and G29.96$-$0.02 the integration time per source was about 14 minutes; for G23.01$-$0.41, G23.71$-$0.20, and IRAS$\,$18566+0408 was about 28 minutes each during epoch 4. The integration time on IRAS$\,$18566+0408 in epoch 1 was approximately 9 minutes. To derive pointing and focus corrections we observed the GBT calibrators: J2148+6107 in epochs 1, 2, 3, 5, and 6; 3C48 in epochs 1, 2, 4, 5, and 6; and J1851+0035 in epochs 1 and 4. The pointing corrections were smaller than 4\arcsec. The system temperature for the six epochs of observations was approximately 23$\,$K. We measured a continuum flux density of 5.95$\,$Jy at 6$\,$cm for 3C48 in epoch 5, which agrees within 9$\%$ with the expected value of 5.45$\,$Jy \citep{2013ApJS..204...19P}. The percentage error is less than 9$\%$ for all other epochs.

\subsubsection{Ku-Band Observations}

The HPBW of the GBT is between 60\arcsec~and 51\arcsec~at 12.2 to 14.8$\,$GHz. The Ku-Band observations were done in beam switching (nod) mode; the two beams are separated by 330\arcsec. We conducted observations in epochs 1 and 5 (see Table \ref{tab:tbl-1}).

\begin{itemize}

\item Epoch 1: The total integration time on-source was 8 min. The system temperature was approximately 30$\,$K. We obtained pointing and focus corrections from observations of J2148+6107; the pointing corrections were smaller than 4\arcsec. We applied atmospheric opacity corrections (at zenith) between 0.011 and 0.014 Nepers within the 12.1 to 14.8$\,$GHz frequency range, which were determined based on data from nearby weather stations accessed through the GBT software CLEO. We also observed the quasar 3C48 to estimate the flux density calibration uncertainty of our observations. We measured S$_\nu$= 1.79\,Jy at 14.488$\,$GHz, which agrees within 5$\%$ with the expected value of 1.88$\,$Jy \citep{2013ApJS..204...19P}.

\item Epoch 5: The total on-source integration time was 4 minutes. The system temperature was approximately 27$\,$K. We applied atmospheric opacity corrections (at zenith) between 0.018 and 0.025 Nepers within the 12.1 to 14.8$\,$GHz frequency range. 

\end{itemize}

\subsubsection{K-Band Observations}

The K-Band observations were conducted in beam switching mode (feed separation $\sim$3\arcmin), dual circular polarization. The GBT HPBW between 19.9 and 22.2$\,$GHz is between 38\arcsec~and 34\arcsec.  We conducted observations in epochs 1, 2, and 5 (Table \ref{tab:tbl-1}). The total integration times on-source were 15, 4, and 6 minutes, respectively. The system temperature varied between 24 and 50$\,$K. We observed the quasars J2148+6107, 3C48, and 3C295 to derive pointing and focus corrections. The pointing corrections were smaller than 6\arcsec. 

\begin{itemize}

\item Epoch 1: we observed 3C295 to check the flux density calibration. We measured a continuum flux density of 1.01$\,$Jy at 22.2$\,$GHz, which agrees within 4$\%$ with the expected value of 0.97$\,$Jy \citep{2013ApJS..204...19P}. The atmospheric opacities at zenith were between 0.037 and 0.071 Nepers (19.9 to 22.2$\,$GHz).

\item Epoch 2: we observed two calibrators (3C48 and J2148+6107). We measured a continuum flux density of 1.35$\,$Jy at 22.2$\,$GHz for 3C48, which agrees within 7$\%$ with the expected value of 1.26$\,$Jy \citep{2013ApJS..204...19P}. The atmospheric opacities at zenith were between 0.023 and 0.039 Nepers (19.9 to 22.2$\,$GHz). 

\item Epoch 5: The atmospheric opacities at zenith were between 0.042 and 0.093 Nepers (19.9 to 22.2$\,$GHz). 

\end{itemize}

\subsection{Westerbork Synthesis Radio Telescope Observations}

We observed the 6$\,$cm H$_2$CO maser in NGC~7538~IRS~1 using the Westerbork Synthesis Radio Telescope\footnote{The WSRT is operated by ASTRON (Netherlands Institute for Radio Astronomy) with support from the Netherlands Foundation for Scientific Research (NWO).} at epochs 7 and 8: 2014 December 07 and 2015 May 29. Only a subset of the array was available for the observations --- eight antennas for the first epoch and six antennas for the second --- with baseline lengths in the range 50--2550$\,$m corresponding to a maximum angular sensitivity of 230\arcsec. For both epochs the maser target was observed for 1.0$\,$hr preceded and followed by observations of 3C48. The data were reduced using AIPS of the NRAO. For both epochs, 3C48 was observed to have a flux density of 5.1$\,$Jy which is 6\% below the expected value of 5.45$\,$Jy \citep{2013ApJS..204...19P}. Aside from initial amplitude and phase calibrations using 3C48, no self-calibration was applied. The spectra were formed as 1024 channels across a 312.5$\,$kHz band for a velocity spacing of 19$\,$m$\,$s$^{-1}$. The images from the two epochs have beams of approximately $60$\arcsec $ \times 3$\arcsec~at position angles of 30 and 10 degrees, respectively. The noise level in a single-channel image is 140$\,$mJy$\,$beam$^{-1}$, consistent with instrumental expectations.

\subsection{Karl G. Jansky Very Large Array Observations}

The 6$\,$cm H$_2$CO maser was also observed with the NRAO Karl G. Jansky Very Large Array (VLA) on June 28, 2015 (epoch 9) in A-configuration (synthesized beam: 0.89$\arcsec ~ \times 0.33\arcsec$, $-65\degr$ P.A.; 8.9$\arcsec$~largest recoverable angular scale). The 6$\,$cm H$_2$CO line ($\nu_0 = 4829.6594\,$MHz, see Table \ref{tab:tbl-1}) was observed with 4096 channels, 4$\,$MHz (248\kms) bandwidth, and a channel width of 0.977$\,$kHz (0.06\kms). We used 3C286 (J1331+305) as flux density and bandpass calibrator and J2230+6946 as complex gain calibrator, which was observed for approximately 3$\,$min before and after NGC$\,$7538$\,$IRS$\,$1. The time on NGC$\,$7538$\,$IRS$\,$1 was $\approx$ 13$\,$min. All data reduction was done using the NRAO data reduction package CASA following standard spectral line data reduction procedures. Assuming a flux density of 7.365$\,$Jy for 3C286 \citep{2013ApJS..204...19P} we measure a flux density of $1.00 \pm 0.01\,$Jy for J2230+6946 using the CASA task {\tt fluxscale}, which agrees within 10\% with the value of 1.10$\,$Jy at C-band listed in the VLA Calibration Manual\footnote{http://www.aoc.nrao.edu/$\sim$gtaylor/csource.html, accessed on 2016 July 14$^{th}$.}.

\section{Results}

\subsection{H$_2$CO, H$_2^{\,13\!}$CO, and H$_2$C$^{\,18\!}$O}

\begin{figure}
\includegraphics{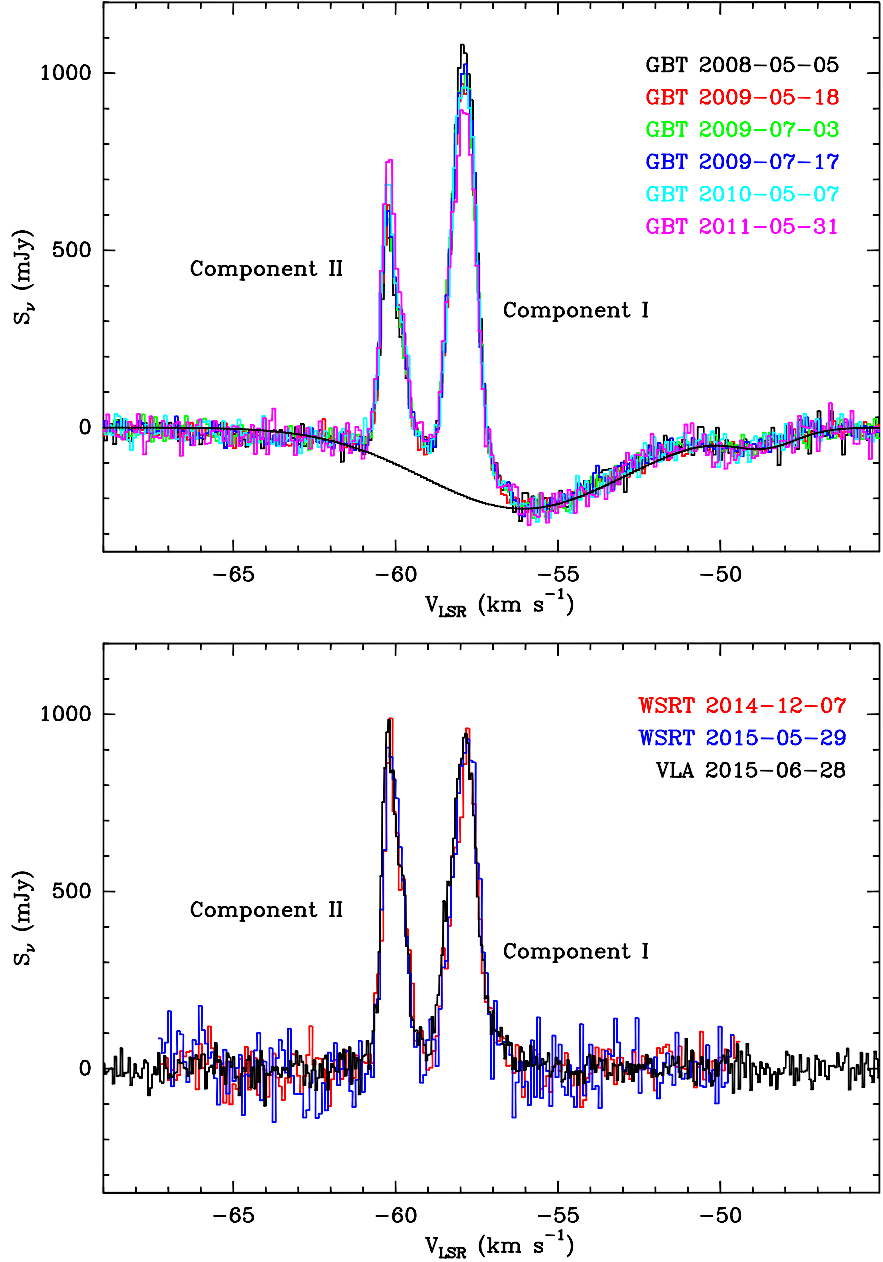}
\vspace*{19.5cm}
\caption{Spectra of the 6$\,$cm formaldehyde masers obtained towards NGC$\,$7538$\,$IRS$\,$1 in the nine observing epochs. {\it Top panel}: the 2008$-$2011 GBT observations. The fit to the averaged formaldehyde double peaked absorption is shown. The consistency in the absorption profiles detected with the GBT (which are not expected to change because of their extended angular size) shows that the observed variability of the masers is not caused by calibration errors. {\it Lower panel}: the 2014 and 2015 observations were obtained with the WSRT and VLA (no formaldehyde absorption was detected within the solid angle of the masers).\label{fig:f1}}
\end{figure}

\begin{figure}
\includegraphics{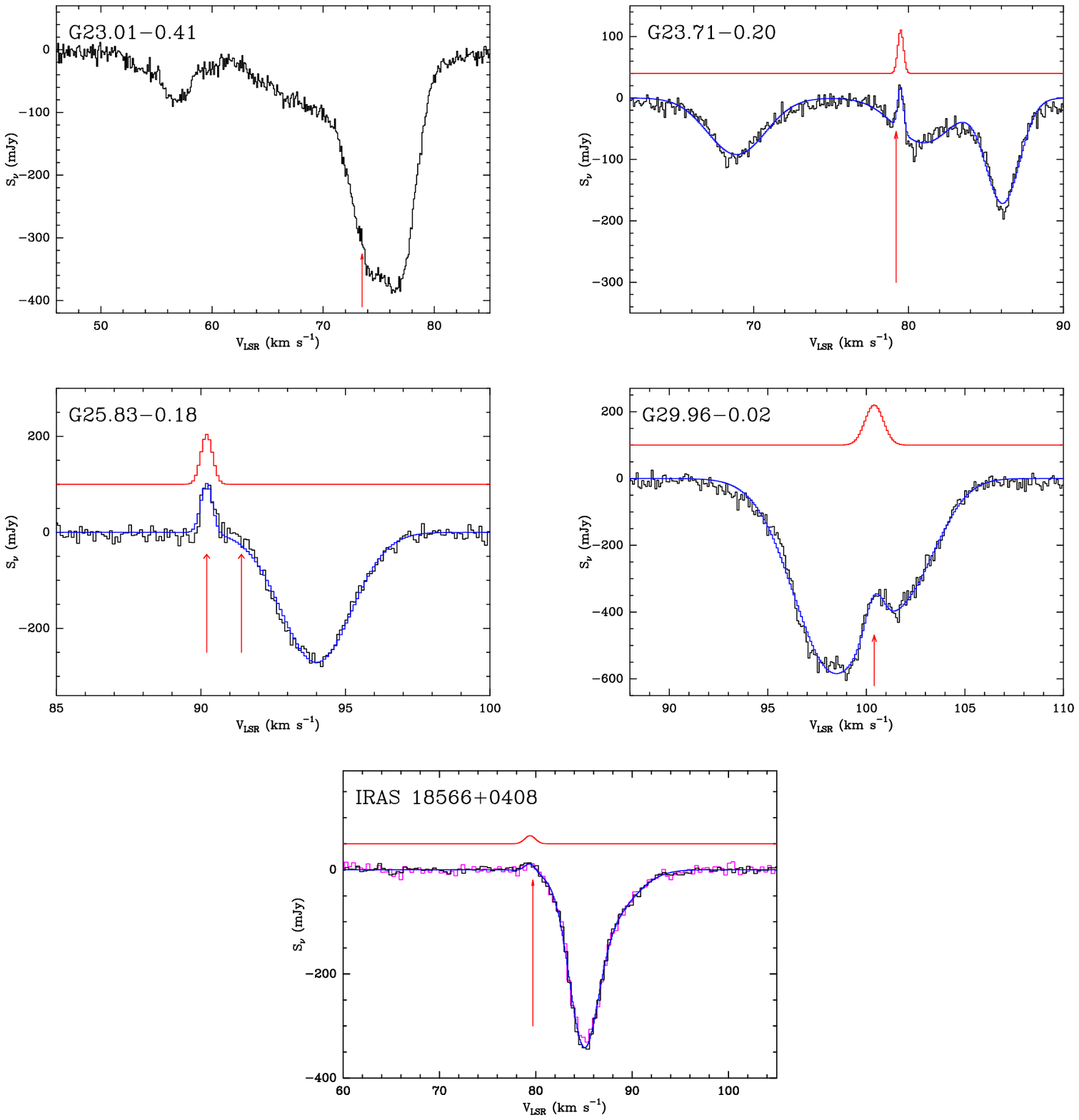}
\vspace*{19.5cm}\caption{Spectra of 6$\,$cm formaldehyde masers toward the other five sources in our sample. All sources were observed on July 17,  2009  (black spectra, epoch 4); IRAS$\,$18566+0408 was also observed in May 5, 2008 (epoch 1; pink spectrum). Red arrows indicate the velocity of previously reported 6$\,$cm formaldehyde masers in the regions \citep{2008ApJS..178..330A, 2004ApJS..154..579A, 1994ApJ...430L.129P}. The blue curves show Gaussian fits to the emission and absorption profiles (a reliable fit of the maser was not possible in the case of G23.01$-$0.41). The red curves show the Gaussian fit of the maser line (the red fits were offset vertically for clarity).\label{fig:f2} }
\end{figure}

\begin{figure}
\includegraphics{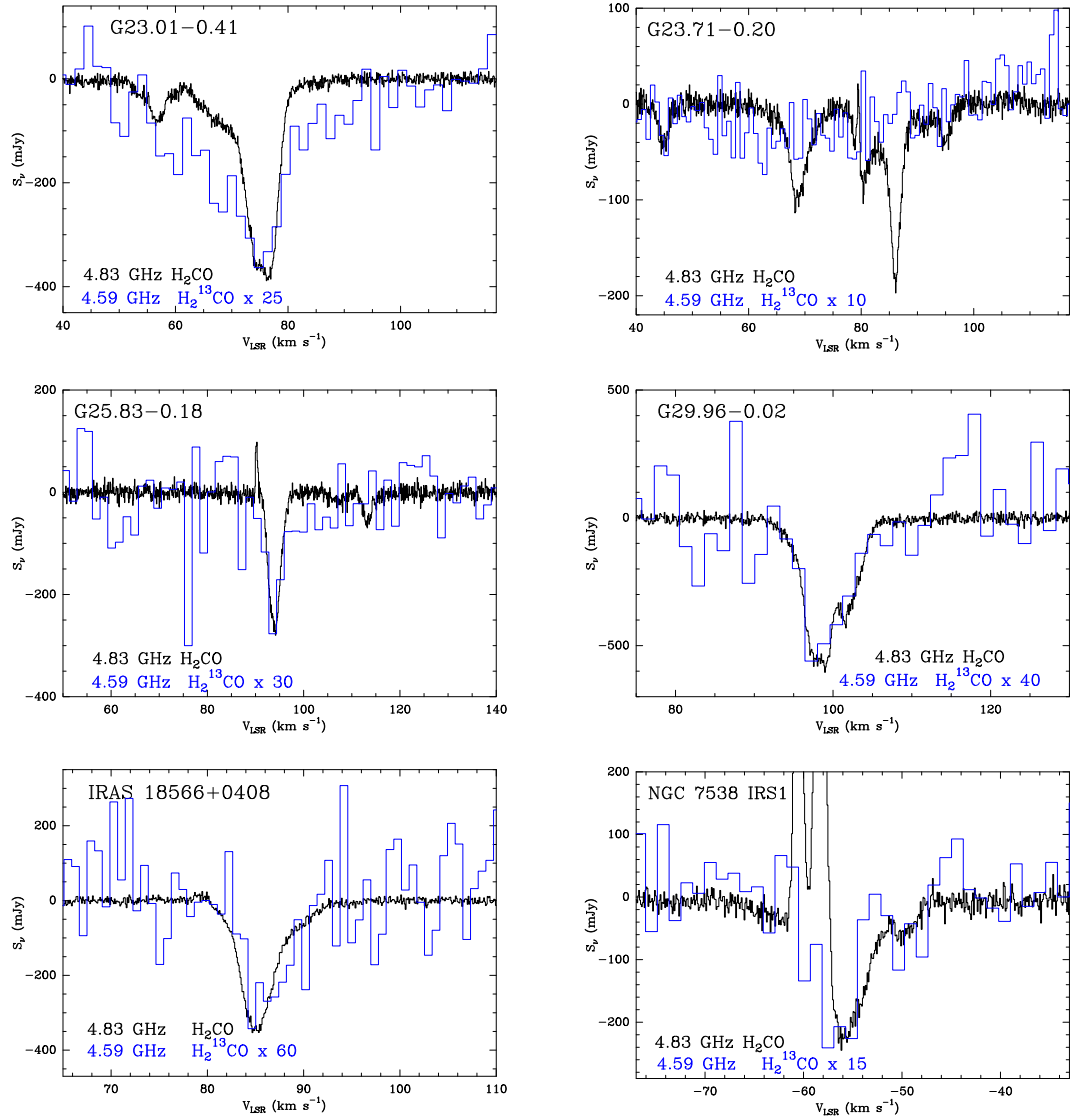}
\vspace*{19.5cm}
\caption{H$_2^{13}$CO isotopologue spectra (4.59$\,$GHz, blue) obtained toward the six sources in our sample overlapped with the spectra of the main isotopologue (4.83$\,$GHz, black). The H$_2^{13}$CO spectra were multiplied by an arbitrary factor to facilitate comparison with the main isotopologue. The isotopologue spectra were smoothed to different channel widths (see Table~\ref{tab:tbl-1}).\label{fig:f3}}
\end{figure}

We detected the double peak 6$\,$cm formaldehyde maser in NGC$\,$7538$\,$IRS$\,$1 in all nine epochs (six GBT observations, two WSRT, and one VLA run; Figure \ref{fig:f1}). In epoch 4 (July 17, 2009), we also observed five other HMSFRs in C-Band. The results of the observations are summarized in Table \ref{tab:tbl-3}: column 1 lists the sources, columns 2 to 5 give the rms and line parameters (peak flux density, peak velocity, and FWHM), and column 6 gives the epochs of observation. The parameters of the absorption lines are presented in Table 4 (the absorption lines were fit simultaneously with the emission lines when detected, see Figure \ref{fig:f2}). We also detected the 1$_{10}$ $-$ 1$_{11}$ (4.59309$\,$GHz) transition of the H$_2^{13}$CO isotopologue in absorption, which is the equivalent quantum transition of the main 6$\,$cm formaldehyde line (previous detections of this transition toward other regions are reported in e.g., \citealt{1976A&A....51..303W}; \citealt{1982A&A...109..344H}; \citealt{1982ApJ...254..538K}). In the case of NGC$\,$7538$\,$IRS$\,$1\footnote{We note that all the absorption lines detected in the NGC$\,$7538$\,$IRS$\,$1 pointing are likely tracing extended clouds and originate from absorption against all continuum sources within the GBT beam, not only the IRS$\,$1 region.}, we found a hint of absorption in all epochs; the average spectra and line parameters are presented in Figure \ref{fig:f3} and Table \ref{tab:tbl-5}. Weak 4.59$\,$GHz H$_2^{13}$CO isotopologue absorption was also detected toward G23.01$-$0.41, G25.83$-$0.18, G29.96$-$0.02, and IRAS$\,$18566+0408 (Table \ref{tab:tbl-5}). H$_2^{13}$CO and H$_2$CO absorption is seen at the same LSR velocities, which suggests similar gas distribution. No isotopologue absorption was detected in G23.71$-$0.20 (Figure \ref{fig:f3}). No 4.59$\,$GHz H$_2^{13}$CO emission lines were detected. Emission overlapped with absorption of the 14.5$\,$GHz H$_2$CO transition was detected in NGC$\,$7538$\,$IRS$\,$1 and it will be discussed in a later work. We also report non-detections of the 13.78$\,$GHz H$_2^{13}$CO and 4.39$\,$GHz H$_2$C$^{18}$O transitions toward NGC$\,$7538$\,$IRS$\,$1 (see Table~\ref{tab:tbl-5}).

\begin{deluxetable}{lclccl}
\tabletypesize{\scriptsize}
\tablecaption{ 6$\,$cm Formaldehyde Masers  \label{tab:tbl-3}}
\tablewidth{0pt}
\tablehead{
\colhead{Source}&  \colhead{rms} & \colhead{S$_{\nu}$} & \colhead{V$_{LSR}$} & \colhead{FWHM} & \colhead{Epoch/Date}\\
\colhead{ } & \colhead{(mJy)} & \colhead{(mJy)} & \colhead{(\kms) } & \colhead{(\kms)}& \colhead{ }}
\startdata
G23.01$-$0.41          &  7 &  \nodata$^1$   & \nodata        & \nodata    & 4/2009-July-17 \\
G23.71$-$0.20$^2$      &  8 &   71(8)  &    79.49(0.02) & 0.43(0.05) & 4/2009-July-17 \\
G25.83$-$0.18$^2$      & 10 &  104(8)  &    90.24(0.02) & 0.48(0.04) & 4/2009-July-17 \\	
G29.96$-$0.02$^2$      & 14 &  120(20) &   100.40(0.05) &  1.1(0.2)  & 4/2009-July-17 \\
IRAS$\,$18566+0408$^2$ &  5 &   16(6)  &    79.6(0.2)   &  1.3(0.6)  & 1/2008-May-05  \\
                       &  6 &   15(3)  &    79.2(0.1)   &  1.3(0.3)  & 4/2009-July-17 \\
NGC$\,$7538$\,$IRS$\,$1$^3$ & 20 & 1295(20) & $-$57.95(0.09) & 0.85(0.19) & 1/2008-May-05  \\ 
                       &    &  661(20) & $-$60.22(0.09) & 0.85(0.19) &                \\ 
                       & 16 & 1184(16) & $-$57.95(0.09) & 0.95(0.19) & 2/2009-May-18  \\ 
                       &    &  753(16) & $-$60.22(0.09) & 0.76(0.19) &                \\ 
                       & 16 & 1215(16) & $-$57.83(0.09) & 0.95(0.19) & 3/2009-July-03 \\ 
                       &    &  715(16) & $-$60.20(0.09) & 0.76(0.19) &                \\ 
                       & 16 & 1242(16) & $-$57.85(0.09) & 0.95(0.19) & 4/2009-July-17 \\ 
                       &    &  736(16) & $-$60.22(0.09) & 0.76(0.19) &                \\ 
                       & 25 & 1177(25) & $-$57.81(0.09) & 0.95(0.19) & 5/2010-May-07  \\ 
                       &    &  812(25) & $-$60.18(0.09) & 0.76(0.19) &                \\ 
                       & 28 & 1109(28) & $-$57.97(0.09) & 1.04(0.19) & 6/2011-May-31  \\ 
                       &    &  884(28) & $-$60.15(0.09) & 0.76(0.19) &                \\ 
                       & 40 &  959(40) & $-$57.78(0.09) & 0.95(0.19) & 7/2014-December-07 \\ 
                       &    &  988(40) & $-$60.14(0.09) & 0.47(0.19) &                    \\ 
                       & 67 &  934(67) & $-$57.78(0.09) & 0.95(0.19) & 8/2015-May-29 \\ 
                       &    &  911(67) & $-$60.24(0.09) & 0.76(0.19) &                \\ 
                       & 30 &  945(30) & $-$57.84(0.06) & 1.03(0.12) & 9/2015-June-28 \\ 
                       &    &  985(30) & $-$60.20(0.06) & 0.73(0.12) &                \\ 
\enddata
\tablenotetext{1}{Emission was blended with absorption and no reliable measurement was possible (see Figure \ref{fig:f2}).}
\tablenotetext{2}{Parameters were obtained from Gaussian fits (Figure \ref{fig:f2}). 1$\sigma$ statistical errors from the fits are listed as uncertainty.}
\tablenotetext{3}{The masers in NGC$\,$7538$\,$IRS$\,$1 deviated significantly from Gaussian profiles, thus, we report the peak channel flux density (spectrum rms reported as uncertainty), peak channel velocity (channel width listed as uncertainty), and full width at half maximum (two times the channel width reported as uncertainty). In the case of the GBT data (observing epochs 1 to 6), the H$_2$CO absorption given in Table \ref{tab:tbl-4} was subtracted from each spectrum before measuring the line parameters of the masers.}
\end{deluxetable}

\begin{deluxetable}{lclcl}
\tabletypesize{\scriptsize}
\tablecaption{ 6$\,$cm Formaldehyde Absorption \label{tab:tbl-4}}
\tablewidth{0pt}
\tablehead{
\colhead{Source} &   \colhead{rms} & \colhead{S$_{\nu}$} & \colhead{V$_{LSR}$} & \colhead{FWHM}\\
\colhead{ } & \colhead{(mJy)} & \colhead{(mJy)} & \colhead{(\kms) } & \colhead{(\kms)}}
\startdata
G23.01$-$0.41          &  7  &  $-$21(1)   &   4.9(0.1)     &  5.2(0.3)  \\    
                            &   &  $-$45(8)   &   56.2(0.3)    &  6.4(0.7)  \\
                   &   &  $-$39(8)   &   56.9(0.1)    &  2.1(0.4)  \\
                   &   &  $-$90(4)   &   69.3(0.7)    & 10.0(1.3)  \\
                   &   &  $-$370(20) &   75.71(0.05)  &  5.6(0.1)  \\                   
G23.71$-$0.20   & 8  &  $-$43(3)   &   44.84(0.06)  &  2.2(0.2)  \\
            &  &  $-$92(2)   &   68.89(0.05)  &  4.4(0.1)  \\
                        &   &  $-$73(3)   &   80.98(0.09)  &  4.4(0.2)  \\
                      &   &  $-$170(3)  &   86.09(0.03)  & 2.57(0.07) \\
                     &   &  $-$21(2)   &   91.0(0.3)    &  4.9(1.3)  \\
                     &   &  $-$36(4)   &   95.1(0.1)    &  2.0(0.3)  \\                                    
G25.83$-$0.18      &  11 &  $-$42(3)   &   11.11(0.07)  &  1.9(0.2)  \\ 
                   & &  $-$271(3)  &   93.97(0.02)  & 2.86(0.04) \\    
                   &   &  $-$19(2)   &   107.2(0.2)   &  3.6(0.5)  \\  
              &   &  $-$57(3)   &   113.29(0.06) &  2.3(0.1)  \\  
G29.96$-$0.02     &  14 &  $-$580(10) &   6.92(0.02)   & 1.23(0.04) \\
                   &   &  $-$170(10) &   8.50(0.07)   &  1.3(0.2)  \\
                   &   &  $-$65(5)   &   49.95(0.07)  &  2.0(0.2)  \\
                    &   &  $-$19(3)   &   57.7(0.4)    &  4.2(0.8)  \\
                    &   &  $-$173(4)  &   66.95(0.03)  & 2.46(0.07) \\  
         &  &  $-$570(10) &   98.31(0.09)  &  4.7(0.1)  \\
                   &   &  $-$270(10) &   102.2(0.1)   &  3.7(0.2)  \\                      
IRAS$\,$18566+0408$^1$ &  3  &  $-$34(3)   &   21.0(0.1)    &  1.4(0.2)  \\
                  &    &  $-$24(3)   &   23.6(0.1)    &  1.5(0.2)  \\
                  &  & $-$237(17) &   87.1(0.1) &  4.6(0.2)   \\
          &    &  $-$58(16)  &   88.9(0.6)    &  8.7(0.7)  \\
NGC$\,$7538-IRS~1$^2$ & 13 & $-$229(4)& $-$56.83(0.08)&  7.3(0.1)  \\
                  &    &  $-$47 (5)  & $-$49.4 (0.1)  &  2.3 (0.3) \\
\enddata
\tablenotetext{~}{~Line parameters from Gaussian fits; 1$\sigma$ statistical errors from the fits are listed as uncertainty. }
\tablenotetext{1}{Average of two GBT spectra: epochs 1 and 4.}
\tablenotetext{2}{Average of GBT spectra obtained in position switching mode. As mentioned in section 3.1, absorption features in the direction of NGC$\,$7538-IRS~1 of all species reported in this work are likely tracing extended clouds seen in absorption against some or all continuum regions within the GBT beam.}
\end{deluxetable}

\begin{deluxetable}{lccccccl}
\tabletypesize{\scriptsize}
\tablecaption{Line Parameters of Isotopologues \label{tab:tbl-5}}
\tablewidth{0pt}
\tablehead{
\colhead{Source}& \colhead{Molecule} & \colhead{Frequency} &  \colhead{rms} & \colhead{S$_{\nu}$} & \colhead{V$_{LSR}$} & \colhead{FWHM} & \colhead{Epoch/Date}\\
\colhead{ } & \colhead{ } & \colhead{(GHz)}  & \colhead{(mJy)} & \colhead{(mJy)} & 
\colhead{(\kms)} & \colhead{(\kms)}& \colhead{ }}
\startdata
G23.01$-$0.41            & H$_2^{13}$CO    & 4.5930885  & 2 & $-$11(1) &    72.5(0.7) & 21(2)    & 4/2009-July-17 \\
G23.71$-$0.20            & H$_2^{13}$CO    & 4.5930885  & 3 & \nodata  &    \nodata   & \nodata  & 4/2009-July-17 \\  
G25.83$-$0.18            & H$_2^{13}$CO    & 4.5930885  & 2 & $-$8(2)  &    94.0(0.8) & 5(2)     & 4/2009-July-17 \\
G29.96$-$0.02            & H$_2^{13}$CO    & 4.5930885  & 4 & $-$13(2) &    98.9(0.5) & 7(1)     & 4/2009-July-17 \\ 
IRAS$\,$18566+0408       & H$_2^{13}$CO    & 4.5930885  & 3 & $-$6(1)  &    86.2(0.3) & 6(1)     & average$^1$     \\ 
NGC$\,$7538$\,$IRS$\,$1  & $^{13}$CH$_3$OH & 14.7822120 & 3 & $-$12(3) & $-$60.8(0.5) & 5(1)     & 1/2008-May-05  \\
                         &  		  &            & 4 & $-$13(3) & $-$59.2(0.4) & 4(1)     & 5/2010-May-07  \\
                         & H$_2^{13}$CO    & 4.5930885  & 3 & $-$16(3) & $-$56.7(0.3) & 4(1)     & average$^2$     \\
                         & H$_2^{13}$CO    & 13.7788041 & 6 & \nodata  &    \nodata   & \nodata  & average$^1$     \\
                         & H$_2$C$^{18}$O  & 4.3887970  & 7 & \nodata  &    \nodata   & \nodata  & average$^2$     \\
\enddata
\tablenotetext{~}{~Line parameters from Gaussian fits; 1$\sigma$ statistical errors from the fits are listed as uncertainty. }
\tablenotetext{1}{Average of two epochs, see Tables~\ref{tab:tbl-1} and \ref{tab:tbl-2}.}
\tablenotetext{2}{Average of GBT spectra obtained in position switching mode.}
\end{deluxetable}

\subsection{CH$_3$OH and $^{13}$CH$_3$OH}

\begin{figure}
\includegraphics{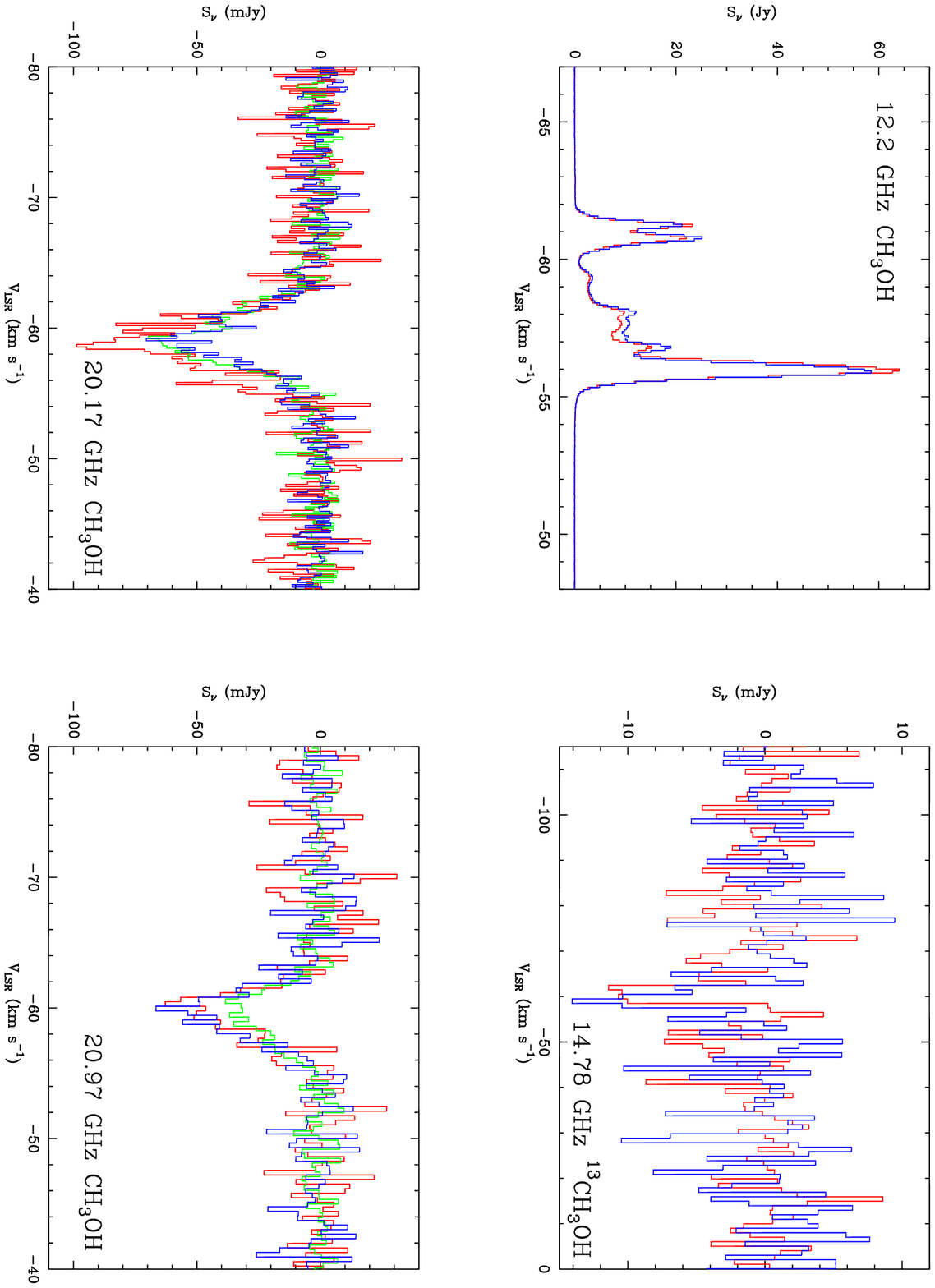}
\vspace*{17.5cm}
\caption{Methanol spectra obtained towards NGC$\,$7538$\,$IRS$\,$1. The spectra of the strong 12.2$\,$GHz methanol masers are shown in the upper left panel, $^{13}$CH$_3$OH isotopologue absorption is shown in the upper right panel, and methanol absorption spectra from two other transitions are shown in the lower panels. Red, green, and blue spectra show data from 2008 (epoch 1), 2009 (epoch 2), and 2010 (epoch 5), respectively (not all lines were observed in all epochs). \label{fig:f4}}
\end{figure}

We observed 12.2$\,$GHz methanol masers in NGC$\,$7538$\,$IRS$\,$1 in 2008 and 2010 (Figure \ref{fig:f4}, Table \ref{tab:tbl-6}). Multiple velocity components were detected, and the flux density of the main peak ($-$56$\,$\kms) slightly decreased between the two epochs. We also detected weak absorption ($-12$ mJy at $-$60 \kms, rms $\sim\,$3$\,$mJy, Table \ref{tab:tbl-5}) of the $^{13}$CH$_3$OH isotopologue (14.782212$\,$GHz) toward NGC$\,$7538$\,$IRS$\,$1 in two epochs (epochs 1 and 5, Ku-Band). This is the equivalent quantum transition of the main methanol isotope at 12.2$\,$GHz. In addition, we detected two K-Band methanol transitions (20.171089$\,$GHz and 20.970651$\,$GHz) in three observing runs (2008, 2009, and 2010, Table \ref{tab:tbl-6}). All methanol lines are shown in Figure \ref{fig:f4}. The 19.9$\,$GHz methanol line was also detected but it will be discussed in a later paper.

~

\subsection{H$_2$O}

\begin{figure}
\includegraphics{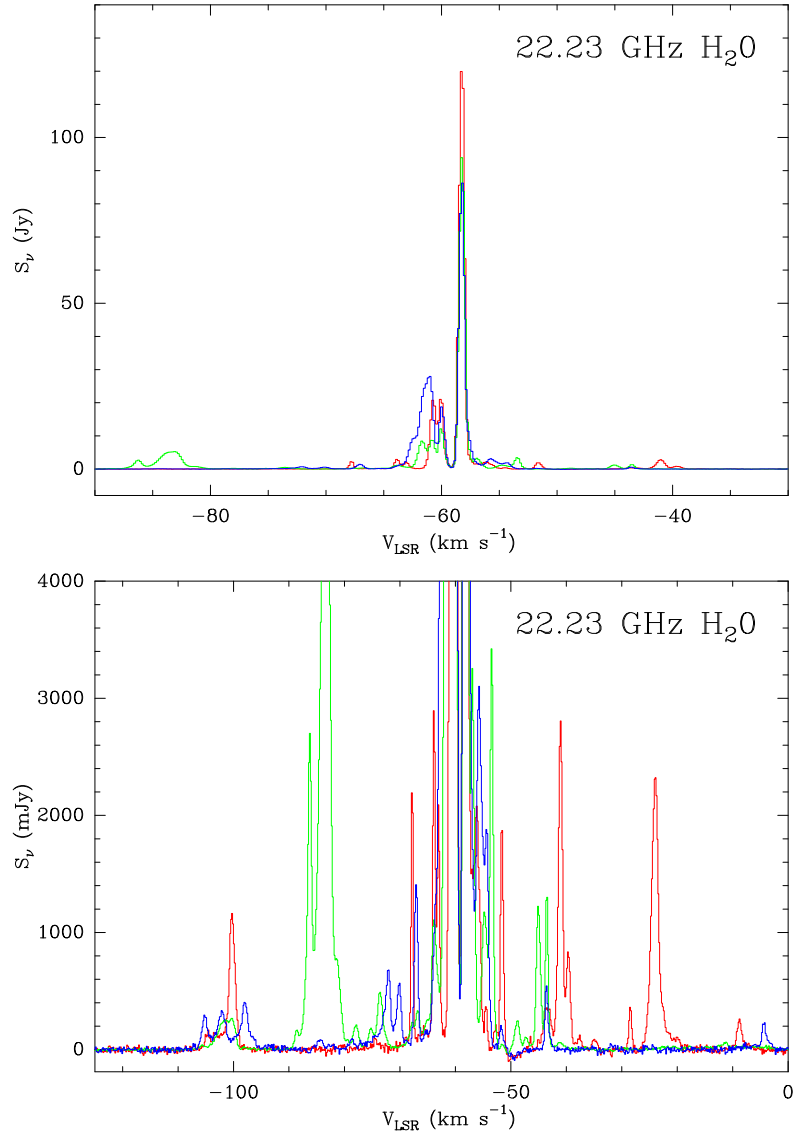}
\vspace*{17.5cm}
\caption{The figure shows H$_2$O spectra towards NGC$\,$7538$\,$IRS$\,$1 from May 2008 (epoch 1; red), May 2009 (epoch 2; green), and May 2010 (epoch 5; blue). The lower panel shows a zoom-in of the weak features (note the change in units to mJy) and displays a broader velocity range than the upper panel. A decrease in the flux density of the main component ($-$58.2\kms) is seen in the upper panel. We also detected water absorption at about $-$50\kms. \label{fig:f5}}
\end{figure}

We conducted observations of the 22.2$\,$GHz water transition in NGC$\,$7538$\,$IRS$\,$1 (2008, 2009, and 2010; Table \ref{tab:tbl-7}, Figure \ref{fig:f5}). Our data show decreasing intensity of the main component ($-$58.2 \kms) and highly variable weaker components (Figure \ref{fig:f5}). We also detected water absorption at LSR velocity of $-$50$\,$\kms, which was mostly masked by a variable water maser in the May 2009 spectrum (green, Figure \ref{fig:f5}). The velocity of this water absorption line is similar to the velocity of a weak 6$\,$cm formaldehyde absorption component (Table \ref{tab:tbl-4}).

\begin{deluxetable}{lccccc}
\tabletypesize{\scriptsize}
\tablecaption{Line Parameters of Methanol Transitions Toward NGC$\,$7538$\,$IRS$\,$1 \label{tab:tbl-6}}
\tablewidth{0pt}
\tablehead{
\colhead{Frequency}  & \colhead{rms} & \colhead{S$_{\nu}$} & 
\colhead{V$_{LSR}$} & \colhead{Linewidth} & 
\colhead{Epoch/Date}\\
\colhead{(GHz)} &\colhead{(mJy)} & \colhead{(Jy)} & 
\colhead{(\kms) } & \colhead{(\kms)} & \colhead{ }}
\startdata
12.1785970 & 15 & 23.247(0.015)   & $-$61.23(0.08) & 1.13(0.15) &1/2008-May-05 \\
         &    & 22.033(0.015)   & $-$60.78(0.08) & 1.05(0.15) & \\
         &    &  3.037(0.015)   & $-$59.35(0.08) & 0.98(0.15) & \\
         &    &  9.797(0.015)   & $-$58.08(0.08) & 1.20(0.15) & \\
         &    &  9.280(0.015)   & $-$57.63(0.08) & 0.68(0.15) & \\
         &    & 15.136(0.015)   & $-$56.80(0.08) & 0.75(0.15) & \\
         &    & 64.089(0.015)   & $-$55.97(0.08) & 2.03(0.15) & \\
         & 21 & 21.181(0.021)   & $-$61.23(0.08) & 0.98(0.15) & 5/2010-May-07 \\
         &    & 25.113(0.021)   & $-$60.78(0.08) & 1.05(0.15) & \\
         &    &  3.517(0.021)   & $-$59.28(0.08) & 0.98(0.15) & \\
         &    & 12.076(0.021)   & $-$58.08(0.08) & 1.13(0.15) & \\
         &    & 10.834(0.021)   & $-$57.63(0.08) & 0.53(0.15) & \\
         &    & 18.992(0.021)   & $-$56.80(0.08) & 0.83(0.15) & \\
         &    & 58.486(0.021)   & $-$55.90(0.08) & 1.95(0.15) & \\
14.782212   & 3 & $-$0.012(0.003)   & $-$60.8(0.5) & 4(1) &1/2008-May-05 \\   
         & 4 & $-$0.013(0.003)   & $-$59.2(0.4) & 3.7(0.9) &5/2010-May-07 \\       
20.1710890 & 13 & $-$0.078(0.004) & $-$58.8(0.1) 	 & 4.7(0.2)   & 1/2008-May-05 \\
         & 5  & $-$0.059(0.001) & $-$59.10(0.06) & 4.7(0.1)   & 2/2009-May-18 \\                        
         &  7 & $-$0.053(0.002) & $-$59.12(0.08) & 4.8(0.2)   & 5/2010-May-07 \\            
20.9706510 & 11 & $-$0.053(0.005) & $-$59.7(0.2)   & 3.9(0.4)   & 1/2008-May-05 \\
         &  4 & $-$0.035(0.002) & $-$59.5(0.1) 	 & 4.6(0.3)   & 2/2009-May-18 \\  
         &  9 & $-$0.054(0.003) & $-$59.6(0.1) 	 & 4.1(0.3)   & 5/2010-May-07 \\ 
\enddata
\tablenotetext{~}{~ The 12.2$\,$GHz CH$_3$OH spectra showed multiple overlapped velocity components which could not be reliably fit using Gaussians, thus, we report the peak flux density (spectrum rms reported as uncertainty), velocity of the peak channel (channel width listed as uncertainty), and the linewidth measured between local minima of overlapping lines (two times the channel width reported as uncertainty). All other transitions were fit with Gaussian functions, 1$\,\sigma$ statistical errors from the fit are shown in parentheses; the linewidth corresponds to the FWHM.}
\end{deluxetable}

\newpage

\subsection{Radio Recombination Lines}

\begin{figure}
\includegraphics{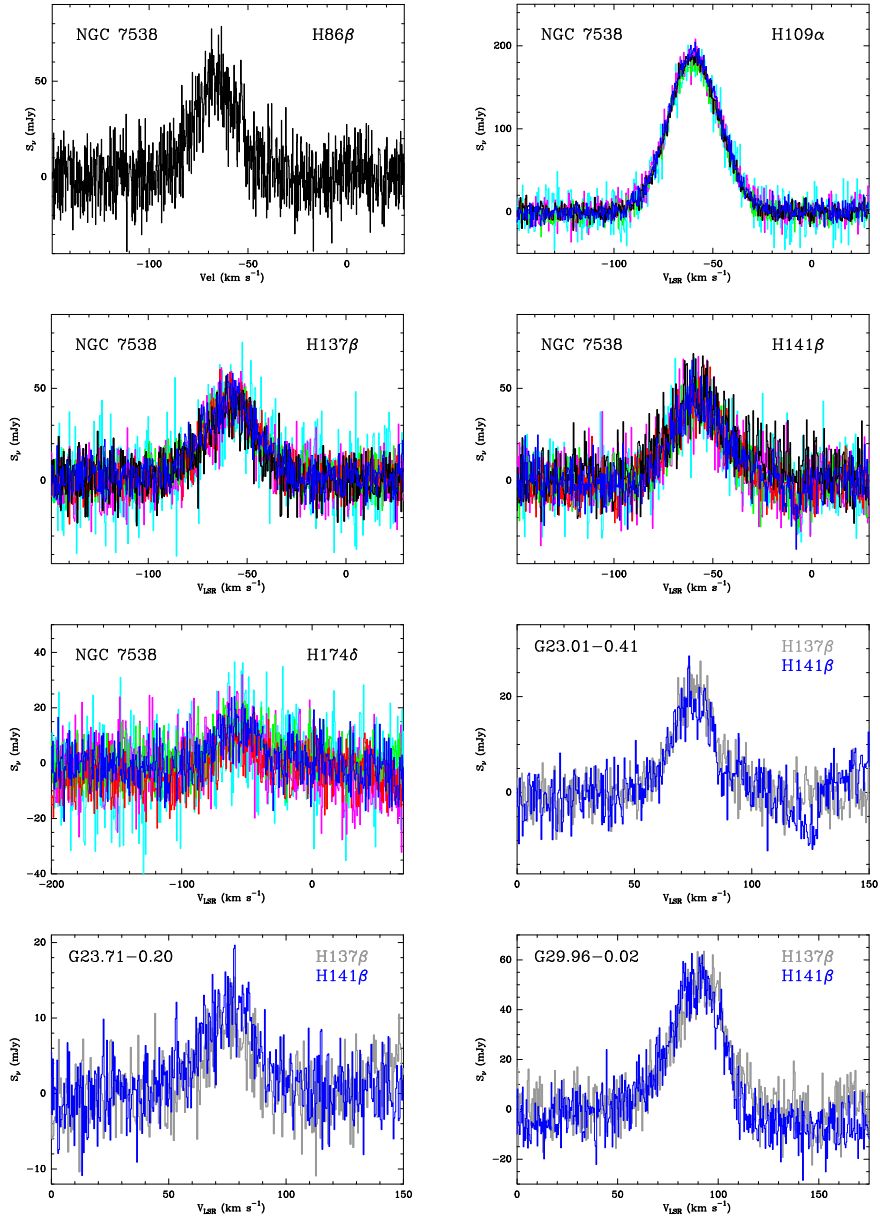}
\vspace*{18.5cm}
\caption{Radio recombination lines detected in this project. The NGC$\,$7538 panels show specific radio recombination line transitions, and different colors represent different observing epochs: May 2008 (epoch 1; black), May 2009 (epoch 2; red), July 2009 (epoch 3; green), July 2009 (epoch 4; blue), May 2010 (epoch 5; pink), and May 2011 (epoch 6; magenta). Not all lines were observed in all epochs (see Table \ref{tab:tbl-1}). The panels of G23.01$-$0.41, G23.71$-$0.20, and G29.96$-$0.02 show two different recombination line transitions (H137$\beta$ in grey; H141$\beta$ in blue) that were observed in epoch 4 (July 17, 2009).\label{fig:f6}}
\end{figure}

We observed four RRLs in C-Band (H109$\alpha$, H137$\beta$, H141$\beta$, and H174$\delta$) and one in K-Band (H86$\beta$) toward NGC$\,$7538$\,$IRS$\,$1 (Figure \ref{fig:f6}, Table \ref{tab:tbl-8}). The HPBW of our C- and K-band observations is large enough to not only include emission from NGC$\,$7538$\,$IRS$\,$1, but also from the ultracompact H{\small II} regions labeled A and C by \citet{1989ApJS...69..831W} (also known as IRS$\,$2 and IRS$\,$3, e.g., \citealt{2004ApJ...605..285S}), and the compact H{\small II} region Sharpless 158 (e.g., \citealt{2016ApJ...824..125L}). This is evident because the RRLs we detected are narrower than the RRLs in NGC$\,$7538$\,$IRS$\,$1 (e.g., \citealt{1995ApJ...438..776G}; \citealt{2004ApJ...605..285S}, \citealt{2008ApJ...672..423K}). We also report detection of H137$\beta$ and H141$\beta$ toward G23.01$-$0.41, G23.71$-$0.20, and G29.96$-$0.02 (Figure \ref{fig:f6}). In all cases, we measured full width at half maximum (FWHM) values between 20 and 35\kms~and found no significant flux density variability within the rms of the observations (Table \ref{tab:tbl-8}, Figure \ref{fig:f6}). The consistency between different epochs exemplifies the reliability of GBT observations for multi-year monitoring studies of spectral lines.

\clearpage

\startlongtable
\begin{deluxetable}{lllc}
\tabletypesize{\scriptsize}
\tablecaption{Line Parameters of Water Masers Toward NGC$\,$7538$\,$IRS$\,$1\label{tab:tbl-7}}
\tablewidth{0pt}
\tablehead{
\colhead{S$_{\nu}$} & \colhead{V$_{LSR}$} & \colhead{FWHM}  & \colhead{Epoch/Date}\\
\colhead{(Jy)} & \colhead{(\kms) } & \colhead{(\kms)} &  \colhead{ }}
\startdata
0.14(0.02)     & $-$104.69(0.08) & 1.0(0.2)    & 1/2008-May-05  \\  
0.14(0.02)     & $-$103.3(0.1)   & 1.0(0.4)    & \\  
0.14(0.02)     & $-$102.0(0.2)   & 1.1(0.4)    & \\  
1.14(0.02)     & $-$100.24(0.01) & 1.24(0.03)  & \\  
0.07(0.01)     & $-$74.3(0.1)    & 1.6(0.3)    & \\
0.108(0.005)   & $-$68.86(0.02)  & 0.86(0.06)  & \\  
2.33(0.07)     & $-$67.77(0.01)  & 0.43(0.02)  & \\  
2.79(0.06)     & $-$63.84(0.01)  & 0.51(0.01)  & \\  
1.88(0.07)     & $-$63.04(0.01)  & 0.52(0.02)  & \\  
20.7(0.4)      & $-$60.73(0.01)  & 0.59(0.02)  & \\  
21.2(0.5)      & $-$60.01(0.01)  & 0.52(0.02)  & \\  
126(3)         & $-$58.24(0.01)  & 0.56(0.02)  & \\  
1.1(0.2)       & $-$57.14(0.09)  & 0.6(0.2)    & \\  
2.01(0.05)     & $-$56.14(0.04)  & 1.31(0.08)  & \\  
0.39(0.02)     & $-$54.49(0.01)  & 0.59(0.04)  & \\  
0.19(0.02)     & $-$52.65(0.04)  & 0.9(0.1)    & \\  
1.95(0.05)     & $-$51.61(0.01)  & 0.61(0.02)  & \\
$-$0.08(0.04)  & $-$49.5(0.5) 	 & 2(1)        & \\
0.11(0.01)     & $-$46.46(0.03)  & 0.38(0.06)  & \\  
0.36(0.03)     & $-$43.35(0.07)  & 1.8(0.2)    & \\  
2.74(0.04)     & $-$41.03(0.01)  & 0.89(0.01)  & \\  
0.82(0.03)     & $-$39.67(0.02)  & 0.98(0.04)  & \\  
0.093(0.007)   & $-$37.71(0.05)  & 1.3(0.1)    & \\  
0.088(0.009)   & $-$34.89(0.05)  & 1.0(0.1)    & \\  
0.37(0.02)     & $-$28.45(0.01)  & 0.57(0.03)  & \\  
2.28(0.04)     & $-$24.01(0.02)  & 1.42(0.04)  & \\  
0.21(0.04)     & $-$22.1(0.2)    & 1.4(0.5)    & \\  
0.245(0.008)   & $-$8.79(0.02)   & 1.02(0.05)  & \\  
0.043(0.007)   & $-$7.2(0.2)     & 1.4(0.4)    & \\
0.033(0.002)   & $-$105.51(0.06) & 1.7(0.2)    & 2/2009-May-18 \\  
0.245(0.004)   & $-$101.62(0.06) & 2.9(0.1)    & \\  
0.14(0.01)     & $-$100.10(0.02) & 1.06(0.09)  & \\   
0.165(0.008)   & $-$88.49(0.03)  & 1.09(0.08)  & \\   
2.5(0.1)       & $-$86.27(0.03)  & 0.96(0.06)  & \\  
5.44(0.09)     & $-$83.45(0.02)  & 2.09(0.05)  & \\   
0.6(0.1)       & $-$81.0(0.1)    & 0.9(0.3)    & \\   
0.086(0.004)   & $-$79.55(0.04) & 1.0(0.1)    & \\   
0.209(0.004)   & $-$77.92(0.01)  & 1.18(0.03)  & \\   
0.18(0.02)     & $-$75.27(0.05)  & 1.0(0.1)    & \\
0.47(0.01)     & $-$73.54(0.02)  & 1.28(0.05)  & \\  
0.041(0.002)   & $-$69.40(0.06)  & 1.5(0.2)    & \\   
0.220(0.007)   & $-$67.64(0.08)  & 1.7(0.2)    & \\ 
0.32(0.01)     & $-$66.83(0.03)  & 1.4(0.1)    & \\   
0.25(0.02)     & $-$65.1(0.1)    & 1.5(0.3)    & \\
0.8(0.5)       & $-$63.91(0.04)  & 0.7(0.1)    & \\   
0.59(0.1)      & $-$63.2(0.4)    & 1.3(0.9)    & \\ 
8.3(0.3)       & $-$61.66(0.02)  & 0.85(0.05)  & \\ 
8.6(0.3)       & $-$60.80(0.01)  & 0.53(0.04)  & \\
12.0(0.3)      & $-$60.00(0.01)  & 0.59(0.02)  & \\
96.1(0.8)      & $-$58.225(0.003)& 0.580(0.007)& \\
3.2(0.7)       & $-$57.0(0.1)    & 1.1(0.3)    & \\
1.16(0.03)     & $-$54.70(0.02)  & 1.29(0.05)  & \\
3.35(0.04)     & $-$53.43(0.01)  & 0.62(0.01)  & \\
$-$0.04(0.01)  & $-$50.49(0.07)  & 0.5(0.2)    & \\
0.242(0.005)   & $-$48.84(0.01)  & 1.02(0.02)  & \\
1.22(0.02)     & $-$45.02(0.01)  & 0.88(0.01)  & \\
1.35(0.02)     & $-$43.51(0.01)  & 0.56(0.01)  & \\
0.055(0.003)   & $-$41.14(0.05)  & 1.9(0.1)    & \\
0.055(0.003)   & $-$11.38(0.05)  & 1.6(0.1)    & \\ 
0.04(0.02)     & $-$107.13(0.09) & 0.5(0.2)    & 5/2010-May-07 \\
0.27(0.01)     & $-$105.13(0.03) & 1.39(0.06)  & \\	        
0.304(0.008)   & $-$102.17(0.03) & 1.96(0.08)  & \\
0.24(0.02)     & $-$97.94(0.02)  & 0.86(0.08)  & \\
0.18(0.02)     & $-$98.09(0.07)  & 3.3(0.3)    & \\	        
0.083(0.008)   & $-$84.43(0.07)  & 1.3(0.2)    & \\
0.085(0.007)   & $-$78.65(0.05)  & 1.2(0.1)    & \\ 
0.21(0.03)     & $-$72.4(0.3)    & 4.3(0.6)    & \\
0.46(0.03)     & $-$72.13(0.03)  & 0.88(0.08)  & \\
0.46(0.04)     & $-$70.12(0.03)  & 0.84(0.09)  & \\	        
0.05(0.02)     & $-$68.2(0.9)    & 1.6(2)      & \\
1.36(0.04)     & $-$67.06(0.01)  & 0.79(0.03)  & \\	        
1.3(0.2)       & $-$63.4(0.3)    & 1.4(0.5)    & \\	
7.9(0.9)       & $-$62.49(0.03)  & 0.76(0.08)  & \\
22.9(0.7)      & $-$61.50(0.03)  & 0.89(0.09)  & \\	        
20(3)          & $-$60.91(0.02)  & 0.60(0.03)  & \\
18.6(0.2)      & $-$59.97(0.01)  & 0.58(0.01)  & \\
87(2)          & $-$58.25(0.01)  & 0.53(0.01)  & \\	        
6.9(0.7)       & $-$57.6(0.1)    & 0.8(0.2)    & \\ 
2.91(0.06)     & $-$55.69(0.02)  & 1.32(0.06)  & \\
1.68(0.08)     & $-$54.34(0.03)  & 0.88(0.06)  & \\	        
0.18(0.07)     & $-$51.7(0.2)    & 0.9(0.4)    & \\	
$-$0.070(0.008)& $-$49.4(0.1)    & 1.5(0.2)    & \\
0.17(0.02)     & $-$47.5(0.2)    & 0.3(0.3)    & \\	   
0.46(0.03)     & $-$43.61(0.07)  & 0.49(0.02)  & \\	
0.14(0.02)     & $-$43.23(0.08)  & 1.1(0.1)    & \\	
0.035(0.009)   & $-$31.38(0.06)  & 0.4(0.1)    & \\	
0.025(0.009)   & $-$25.7(0.3)    & 1.1(0.7)    & \\	
0.04(0.01)     & $-$24.38(0.06)  & 0.4(0.1)    & \\	
0.05(0.02)     & $-$7.8(0.1)     & 0.6(0.3)    & \\	
0.20(0.02)     & $-$4.30(0.05)   & 1.2(0.1)    & \\
\enddata
\tablenotetext{~}{~Line parameters from Gaussian fits; 1$\,\sigma$ statistical errors from the fit are shown in parentheses. The rms of the spectra are: 19$\,$mJy (1/2008-May-05), 7$\,$mJy (2/2009-May-18), 10$\,$mJy (5/2010-May-07).}
\end{deluxetable}

\begin{deluxetable}{lccccccl}

\tablecaption{Radio Recombination  Lines \label{tab:tbl-8}}
\tablewidth{0pt}
\tablehead{
\colhead{Source}& \colhead{RRL} & \colhead{Frequency} &  \colhead{rms} & \colhead{S$_{\nu}$} & \colhead{V$_{LSR}$} & \colhead{FWHM} &\colhead{Epoch/Date}\\
\colhead{ } & \colhead{ } & \colhead{(GHz)}  & \colhead{(mJy)} &\colhead{(mJy)} & \colhead{(\kms) } & \colhead{(\kms)}& \colhead{ }}
\startdata
G23.01$-$0.41        & H137$\beta$  & 5.00503238 & 4  & 22(1)  & 75.6(0.3)    & 20.8(0.7) & 4/2009-July-17\\	
                     & H141$\beta$  & 4.59384760 & 4  & 19(1)  & 74.5(0.4)    & 20.1(0.9) & \\
G23.71$-$0.20        & H137$\beta$  & 5.00503238 & 2  & 12(1)  & 75.9(0.6)    & 28(2)     & 4/2009-July-17\\
                     & H141$\beta$  & 4.59384760 & 3  & 10(1)  & 75.3(0.7)    & 20(2)     & \\
G29.96$-$0.02        & H137$\beta$  & 5.00503238 & 7  & 53(1)  & 88.5(0.3)    & 27.4(0.8) & 4/2009-July-17\\   
                     & H141$\beta$  & 4.59384760 & 8  & 51(1)  & 91.0(0.4)    & 30.2(0.8) & \\  
NGC$\,$7538          & H86$\beta$   &19.97816269 & 13 & 51(2)  & $-$66.5(0.4) & 26.2(0.9) & 1/2008-May-05  \\  
                     & H109$\alpha$ & 5.00892233 & 7  & 198(3) & $-$59.3(0.2) & 33.9(0.5) & 1/2008-May-05   \\
                      &              &	         & 6  & 187(1) & $-$59.6(0.1) & 30.1(0.2) & 2/2009-May-18  \\                     
                      &              &	         & 7  & 180(1) & $-$59.8(0.1) & 30.2(0.2) & 3/2009-July-03 \\
                     &              &	         & 8  & 192(1) & $-$59.5(0.1) & 30.7(0.2) & 4/2009-July-17 \\
                     &              &	         & 8  & 185(2) & $-$59.7(0.2) & 31.7(0.3) & 5/2010-May-07  \\
                     &              &	         & 17 & 184(3) & $-$59.2(0.2) &	30.5(0.5) & 6/2011-May-31  \\
                     & H137$\beta$  & 5.00503238 & 9  & 39(2)  & $-$58.8(0.6) & 30(1)     & 1/2008-May-05  \\
		     &	            & 	         & 6  & 42(1)  & $-$58.2(0.4) & 31(1)     & 2/2009-May-18  \\
                     &              &	         & 7  & 41(1)  & $-$58.5(0.4) & 31(1)     & 3/2009-July-03 \\
                     &              &	         & 8  & 42(1)  & $-$58.6(0.4) & 32(1)     & 4/2009-July-17 \\
                     &              &            & 8  & 43(2)  & $-$57.6(0.6) & 30(1)     & 5/2010-May-07  \\
                     &              &	         & 16 & 39(2)  & $-$57(1)     & 34(2)     & 6/2011-May-31  \\
                     & H141$\beta$  & 4.59384760 & 10 & 42(2)  & $-$56.4(0.7) &	34(2)     & 1/2008-May-05  \\
                     &              &            &  8 & 46(1)  & $-$58.0(0.5) & 33(1)     & 2/2009-May-18  \\
                     &              &            &  9 & 41(1)  & $-$57.8(0.5) &	30(1)     & 3/2009-July-03 \\
                     &              &            &  7 & 46(1)  & $-$57.3(0.5) & 32(1)     & 4/2009-July-17 \\
                     &              &            & 10 & 43(2)  & $-$58.1(0.7) &	27(2)     & 5/2010-May-07  \\
                     &              &            & 14 & 42(2)  & $-$58.1(0.8) &	32(2)     & 6/2011-May-31  \\
                     & H174$\delta$ & 4.82617096 & 7  & 12(1)  & $-$57(2)     & 37(4)     & 2/2009-May-18  \\
                     &              &	         & 6  & 12(1)  & $-$56(2)     & 34(4)     & 3/2009-July-03 \\
                     &              &	         & 6  & 15(1)  & $-$59(1)     & 36(3)     & 4/2009-July-17 \\
                     &              &	         & 9  & 16(2)  & $-$59(2)     & 33(4)     & 5/2010-May-07  \\
                     &              &	         & 13 & 11(3)  & $-$54(3)     & 25(7)     & 6/2011-May-31  \\
\enddata
\tablenotetext{~}{Radio recombination lines frequencies were obtained from Splataloque. Line parameters from Gaussian fits; 1$\,\sigma$ statistical errors from the fit are shown in parentheses.}
\end{deluxetable}

\clearpage

\section{Discussion}

\subsection{Formaldehyde} 

\subsubsection{H$_2$CO Masers in NGC$\,$7538$\,$IRS$\,$1} 

\begin{figure}
\includegraphics{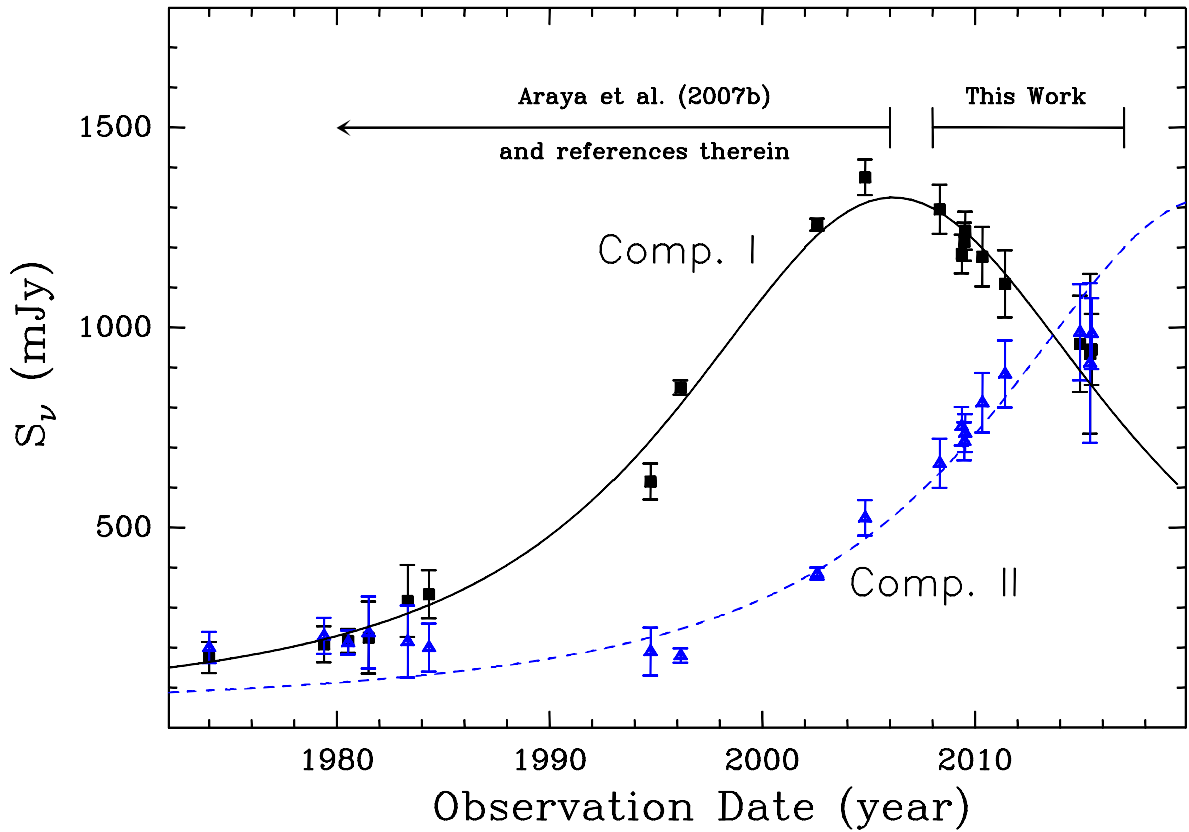}
\vspace*{13.5cm}
\caption{Light curve of the two formaldehyde maser components in NGC$\,$7538$\,$IRS$\,$1; $\pm 3\sigma$ error bars are shown. The black line shows a fit of the Component I data by assuming that the maser is generated by a spinning-ellipsoidal unsaturated maser region (see Section 4.1.1). The blue dashed curve is the same black fit shifted by 15$\,$years to highlight the similar variability behavior of both maser components and the possibility that changes in the line-of-sight amplification path through the maser clouds could be responsible for the observed variability.\label{fig:h2co_masers_light_curve}}
\end{figure}

In contrast to many other astrophysical masers, the two individual 6$\,$cm formaldehyde maser velocity components in NGC$\,$7538$\,$IRS$\,$1 have persisted for four decades. Figure \ref{fig:h2co_masers_light_curve} shows the light-curve of these masers, including the measurements reported in this work and data from the literature. Our data confirm the statement by \citet{2007ApJS..170..152A} that the flux density rate of change of the brighter maser (Component I) decreased approximately from 2004. Specifically, our light curve indicates that Component I reached a maximum flux density sometime near 2004, and its flux density has been decreasing ever since, while the flux density of Component II started increasing after 1996. 

\begin{figure}
\includegraphics{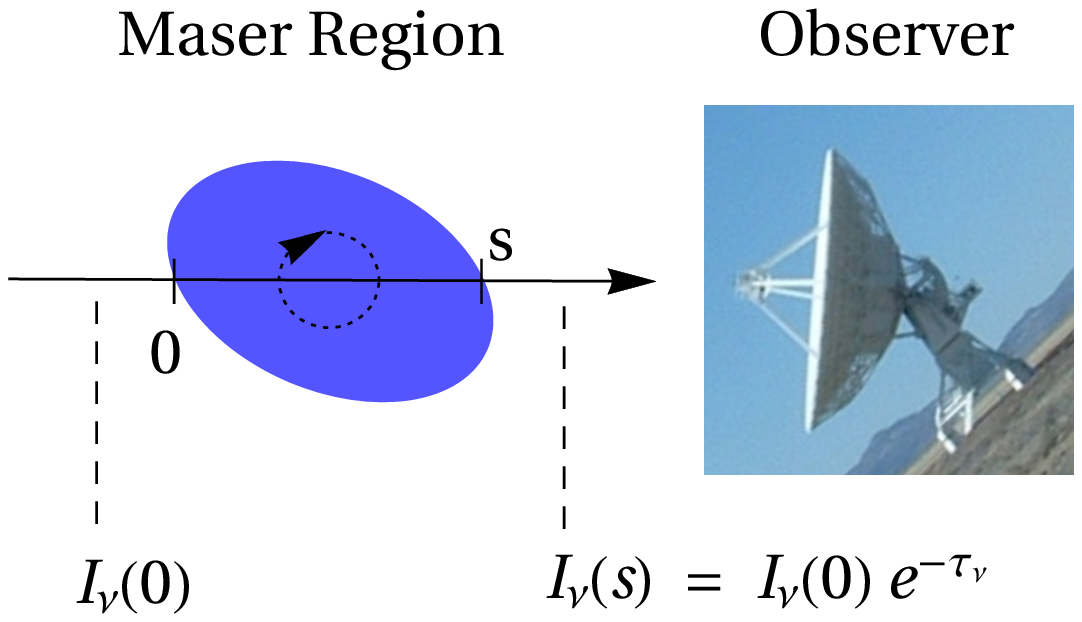}
\vspace*{9.5cm}
\caption{Simple model used to explain the observed variability. As the maser region slowly rotates, a different amplification path ($0 \rightarrow s$) results in a change in the maser gain ($\tau_\nu$), and thus, a change in the observed intensity. \label{fig:model}}
\end{figure}

As noted by \citet{2007ApJS..170..152A}, both maser components showed little variability from their detection in 1974 until the early 1980s, when Component I showed an increase in flux density of the order of 30 mJy$\,$year$^{-1}$, while Component II showed no variability until sometime after 1996, when a similar variability slope was detected (see Figure \ref{fig:h2co_masers_light_curve}). \citet{2007ApJS..170..152A}  mentioned the possibility that some type of perturbation could have passed though one maser region causing the variability, and then reached Component II $\sim$14 years later, triggering a similar variability behavior with a time delay. Our observations show that the similar variability of both maser components has indeed continued (Figure \ref{fig:h2co_masers_light_curve}). However, the hypothesis that some type of propagating phenomenon (e.g., a shock front passing through the medium) is responsible for the variability of both maser components seems now unlikely because one would expect some change in the velocity and/or linewidth of the masers, but our data do not show a significant change in kinematic parameters. Thus, the similar variability may not be caused by a single triggering mechanism that affected both maser regions with a time delay, but rather by similar properties of two independent maser clouds. For instance, the variability could be attributed to a change in the maser cloud physical depth as a function of time (i.e., a change in the amplification/gain path-length through a uniformly pumped medium in our line-of-sight). In a simple amplification model, the maser region can be assumed to have an ellipsoidal shape, spin as a solid body with constant angular momentum (giving velocity coherence along the line-of-sight), and constant pumping. As time passes, the rotation would change the maximum line-of-sight amplification path through the maser cloud $s(t)$:

\begin{equation}
s(t) = \frac{2 \, a \, b}{\sqrt{[a \, sin(\omega \, (t - t_0))]^2 + [b \, cos(\omega \, (t - t_0))]^2 }},
\end{equation}

\noindent where $a$, $b$, $\omega$ are the semimajor axis, semiminor axis, and angular velocity, respectively. This change in amplification path would modify the maser's flux density because the maser gain would change as $\tau_\nu \propto s(t)$ (see Figure \ref{fig:model}).

The black curve in Figure \ref{fig:h2co_masers_light_curve} shows the result of fitting the light-curve of maser Component I to this simple rotating ellipsoidal model. The model assumes unsaturated maser amplification and that the maximum line-of-sight segment across the ellipsoid produces all observed flux density (1D approximation). This is clearly a zero-order approximation because contributions from other lines-of-sight across the ellipsoid are neglected. Nevertheless, it is a useful model to discuss how changes in the maser depth could explain the observed variability. In the context of this model, our data indicate that the ellipsoid was observed near when its major axis was in our line-of-sight. Thus, assuming a prolate geometry, the maximum line-of-sight gain length would dominate the amplification, which justifies the use of the 1D approximation. The model shown in Figure \ref{fig:h2co_masers_light_curve} (black curve) corresponds to a semi-major axis of $\sim 250\,$AU, an aspect ratio between the major and minor axes of $\sim 10$, a rotation period of $\sim 1500\,$years, and its major axis was in the direction of our line-of-sight in 2006. 

We note that the model has a degeneracy in the physical size of the ellipsoid, for instance, the maser cloud could be larger (e.g., $\sim 2000\,$AU$ \times 200\,$AU) if the pumping mechanism were less efficient. \citet{2003ApJ...598.1061H} reported a projected size of Component I of $\approx 200$~AU, and assumed a gain length of $\approx 2000$~AU in their analysis. Thus, the physical dimensions of the ellipsoid used in our model are similar to their observed values. Moreover, \citet{2003ApJ...598.1061H} reported a velocity gradient of 1900$\,$km$\,$s$^{-1}\,$pc$^{-1}$, which would correspond to one complete revolution of the ellipsoid in $\approx 3000$~yr. Thus, the line-of-sight ray through the ellipsoid would exhibit maximal and minimal path lengths every 1500~yr which is similar to the results of our model.

The most interesting aspect of this simple model is that the same curve used to fit the data of Component I also fits well the light-curve of Component II by simply assuming a time-shift of 15$\,$years (blue dashed curve, Figure \ref{fig:h2co_masers_light_curve}).\footnote{However, we note that the flux density of Component II measured in 2014 and 2015 seems lower than expected from the model.} In this interpretation, the similar variability profile of both maser regions is not connected or triggered by an external event with a time delay, but rather is a consequence of both maser regions having similar physical shapes and overall velocity gradients. However, it should be clear that the model is only a simple approximation to the current data. The rotation period of the ellipsoid is very long with respect to the sampling time ($>$ 1,000 years vs 40 years), thus, it is highly unlikely that two unrelated maser clouds are almost perfectly oriented with their major axes in our line-of-sight with less than a few decades difference. Instead, the model simply suggests that the projected line-of-sight depth can be fit by an ellipsoidal morphology over the period of observations.

Recently, \citet{2017MNRAS.466.4364C} report observations of the 6$\,$cm formaldehyde masers in NGC$\,$7538 obtained with the 65$\,$m Shanghai Tianma Radio Telescope (TMRT) in 2016 March and April. They report peak flux densities of 1.314$\,$Jy (Component I) and 1.434$\,$Jy (Component II), i.e., they found that by 2016, the flux density of Component II was greater than Component I, which is qualitatively consistent with the extrapolation of our data as shown in Figure \ref{fig:h2co_masers_light_curve}. However, the flux densities reported by \citet{2017MNRAS.466.4364C} are greater than the expected values from extrapolation of the data reported here. It is unclear if this discrepancy is due to significant deviations from the extrapolated trends, or whether the flux densities reported by \citet{2017MNRAS.466.4364C} were overestimated. Follow up observations within the next 5 to 10 years are required to explore deviations from the long-term variability shown in Figure \ref{fig:h2co_masers_light_curve}, to investigate symmetry in the variability profiles before and after maxima, and similarities between the light-curves of Components I and II. Such data are needed to develop a more realistic 3D model of the maser regions.

\subsubsection{H$_2$CO Masers: Other Sources} 

As mentioned in Section 2.1, in addition to NGC 7538$\,$IRS$\,$1, five other formaldehyde maser sources were observed on 2009 July 17 (Table \ref{tab:tbl-2}). Here we discuss the observations in the context of previous formaldehyde maser detections: 

\textbf{G23.01$-$0.41:} is a region of active high-mass star formation at the distance 4.6$\,$kpc \citep{2009ApJ...693..424B} in the Scutum constellation. The bolometric luminosity of G23.01$-$0.41 is $4\times10^4$$\,$\lsols~\citep{2014A&A...565A..34S}. Clear outflow activity is revealed by $^{12}$CO observations \citep{2008ApJ...673..363F} and by intense $4.5$\,$\mu$m excess emission from a hot molecular core (HMC; \citealt{2008ApJS..178..330A}; \citealt{2009ApJ...702.1615C}). A weak ($\approx 50\,$\mjyb) formaldehyde maser was detected towards the HMC by \citet{2008ApJS..178..330A}.  As shown in Figure \ref{fig:f2}, upper left panel (red arrow), this maser is blended with stronger formaldehyde absorption at $\sim76$$\,$\kms~and weaker and broader one at $\sim 69$$\,$\kms. The presence of formaldehyde absorption in this region is well known (e.g., \citealt{1980A&AS...40..379D}), and a double peak absorption profile has also been seen in OH \citep{2014A&A...565A..34S}. Given the overlapping absorption lines, a reliable measurement of the maser's line parameters is not possible. Based on the asymmetry observed in the spectrum, we estimate a maser peak flux density of the order of 6$\sigma$ (40$\,$mJy), which is similar to the flux density of the maser detected in 2006 \citep{2008ApJS..178..330A}. We conclude that no significant variability of the maser is found between the 2006 and 2010 epochs. 

\citet{2017MNRAS.466.4364C} also report observations of the formaldehyde maser. Even though they did not explicitly discuss variability of this source, \citet{2017MNRAS.466.4364C} list line parameters of the maser in their Table~1 and show the spectrum after removing the absorption in their Figure~1. However, the line they report is excessively broad (1.38\kms~versus 0.4\kms, \citealt{2008ApJS..178..330A}) and it has a different $V_{LSR}$ than expected (75.3\kms~versus 73.6\kms, \citealt{2008ApJS..178..330A}), thus, the emission feature they report could be an artifact caused by overlapping absorption lines.

\textbf{G23.71$-$0.20 (IRAS$\,$18324$-$0820):} is a high-mass star-forming region located at distance of 6.21$\,$kpc \citep {2014ApJ...781..108S}. The maser is located towards an extended radio continuum source \citep{2005AJ....130..586W, 2006ApJ...643L..33A}. No HMC has been detected in this region \citep{2013ApJS..207...12R}. The 6$\,$cm formaldehyde maser overlaps with absorption (Figure \ref{fig:f2}). We measure a peak flux density of 71(8)$\,$mJy at 79.5$\,$\kms, which is similar to the line parameters measured with the VLA in A configuration (\citealt{2006ApJ...643L..33A}; 60$\,$mJy$\,$beam$^{-1}$, 4.5$\,$mJy$\,$beam$^{-1}$ rms, 79.2$\,$\kms, January 2005) but greater than the VLA B-configuration measurement (\citealt{2006ApJ...643L..33A}; 44$\,$mJy$\,$beam$^{-1}$, 6$\,$mJy$\,$beam$^{-1}$ rms, 79.2$\,$\kms, April 2005). Although we cannot rule out low level variability, the formaldehyde maser in G23.71$-$0.20 did not show extreme variability between 2005 and 2009.

\textbf {G25.83$-$0.18 (IRAS$\,$18324$-$0820):} is a massive star-forming region at a distance of 5$\,$kpc \citep{2011MNRAS.417.2500G} where formaldehyde masers were detected at the center of an infrared dark cloud  \citep{2008ApJS..178..330A}. The source is characterized by 6.7$\,$GHz masers and a HMC \citep{2006MNRAS.367..553P}. As reported by \citet{2008ApJS..178..330A}, the maser is found at the blue-shifted wing of a formaldehyde absorption feature (Figure \ref{fig:f2}). We were able to simultaneously fit the main maser feature and the absorption (Figure \ref{fig:f2}). As in the case of NGC$\,$7538$\,$IRS$\,$1 and IRAS$\,$18566+0408, the formaldehyde maser in G25.83$-$0.18 has a double peak profile \citep{2008ApJS..178..330A}. In our spectrum, the weak formaldehyde maser component is significantly blended with the absorption and its line parameters cannot be measured. In the case of the main formaldehyde maser peak, no significant variability was detected with respect to the \citet{2008ApJS..178..330A} observations of 2005. No significant variability is seen with respect to the \citet{2017MNRAS.466.4364C} observations either. 

\textbf {G29.96$-$0.02 (IRAS$\,$18434$-$0242):} is a well known ultra-compact cometary H{\small II} region and HMC at a distance of 5.3 kpc \citep{2014ApJ...783..130R}. The HMC is located $\approx$2.6\arcsec ~to the West of the UCH{\small II} region, and the velocity structure of the HMC is consistent with a rotating massive toroid (\citealt{2011A&A...525A.151B}, \citealt{2013A&A...552A.123B}) with internal substructure \citep{2007A&A...468.1045B}. The formaldehyde maser overlaps with absorption (Figure \ref{fig:f2}). We measured a maser peak flux density of 120(20)$\,$mJy at 100.4$\,$\kms, which is twice the value observed by \citet{2003ApJ...598.1061H}, \citet{1994ApJ...430L.129P}, and \citet{2017MNRAS.466.4364C}. Future interferometric  observations are needed to investigate variability of this maser. 

\textbf {IRAS$\,$18566+0408 (G37.55+0.20):} is the only known region where periodic flares of formaldehyde masers have been detected. The flares are correlated with 6.7$\,$GHz methanol maser flares \citep{2010ApJ...717L.133A} and 6.035$\,$GHz OH flares \citep{2012ApJ...750..170A,2016AAS...22734702H}. IRAS$\,$18566+0408 is located at a kinematic distance of 6.7$\,$kpc \citep{2004ApJS..154..579A} and has bolometric luminosity about $6 \times 10^4$\lsols~\citep{2007A&A...470..269Z}. We detected the formaldehyde maser in May 2008 and July 2009. As shown in Figure \ref{fig:f2}, no flares or significant variability of the maser were detected. The average (two epochs) of the peak flux density is 15$\,$mJy at 79.4(0.2)$\,$\kms. \citet{2017MNRAS.466.4364C} also report detection of the maser, however, the line they found is significantly broader than previous observations (2.98\kms~compared to $< 2$\kms, e.g., \citealt{2007ApJ...654L..95A}). Further observations are needed to confirm changes in linewidth.

\subsubsection{H$_2^{13}$CO Absorption}

As mentioned in Section 2, Figure \ref{fig:f3} shows the H$_2^{13}$CO absorption (blue) scaled up by a multiplicative factor to facilitate comparison with the main isotopologue line (6$\,$cm H$_2$CO). In the case of NGC$\,$7538 (Figure \ref{fig:f3}, bottom right panel), the H$_2^{13}$CO absorption matches the main formaldehyde velocity component at $-56$\kms, and we also find evidence for H$_2^{13}$CO absorption at the velocity of the weaker formaldehyde line at $-50$\kms. However, the weaker H$_2^{13}$CO absorption at $-50$\kms~is in part due to the hyperfine structure of the main absorption at $-56$\kms. Following \citet{1976A&A....51..303W} and \citet{1980A&A....82...41H}, we estimate a column density ratio [H$_2^{12}$CO]/[H$_2^{13}$CO] = $30 \pm 20$ for the lower 6$\,$cm K-doublet energy level\footnote{Hereafter, the [H$_2^{12}$CO]/[H$_2^{13}$CO] refers to the column density ratio of the lower energy K-doublet of the 6$\,$cm transition of both isotopologues.}. This estimate takes into account the hyperfine structure of the H$_2^{13}$CO line (\citealt{1971ApJ...169..429T}, \citealt{1976A&A....51..303W}), and assumes $T_{ex} \sim T_{CMB}$, which implies optically thin absorption (note that \citet{2003ApJ...598.1061H} reported optically thin 6$\,$cm H$_2$CO absorption based on VLA CnB observations, i.e., $\tau = 0.093$). The large uncertainty in our [H$_2^{12}$CO]/[H$_2^{13}$CO] determination is due to the low signal-to-noise of the H$_2^{13}$CO detection (Figure \ref{fig:f3}), the uncertain values of excitation temperature and the photon trapping correction ($f_{12/13}$). We assume that the value of $f_{12/13}$ is between 1.0 and 1.4 (see Figure~4 of \citealt{1980A&A....82...41H}; optically thin absorption), given that the 6$\,$cm H$_2$CO transition is likely to trace molecular gas with densities of less than 10$^{5.5}\,$cm$^{-3}$ as observed in other star forming regions (e.g., \citealt{2011ApJ...736..149G} and \citealt{2015A&A...573A.106G}). We note that based on H$_2$CO observations of the 29$\,$GHz and 48$\,$GHz transitions, \citet{2011ApJ...742...58M} report greater molecular densities ($n(H_2) = 10^{5.78}\,$cm$^{-3}$), however the 6$\,$cm H$_2$CO transition likely traces a larger fraction of the lower density molecular envelope. If the molecular density traced by the 6$\,$cm H$_2$CO is greater than 10$^{5.5}\,$cm$^{-3}$ and/or if the optical depth is greater (e.g., due to clumpiness or greater excitation temperature), then our [H$_2^{12}$CO]/[H$_2^{13}$CO] determination would be a lower limit (e.g., \citealt{1973ApJ...181..781F}; \citealt{1974A&A....36..245W}; \citealt{1977ApJ...214...50L}). Likewise, in the case of the other sources in our sample (Figure \ref{fig:f3}), the [H$_2^{12}$CO]/[H$_2^{13}$CO] values have large uncertainties due to a combination of sensitivity, optical depth effects and unknown filling factors and excitation temperatures. Specifically, we obtain [H$_2^{12}$CO]/[H$_2^{13}$CO] = 49 $\pm$ 33 (G29.96$-$0.02), $>30$ (G23.01$-$0.41), $>60$ (IRAS$\,$18566+0408), and $>40$ (G25.83$-$0.18). These values are similar to previous studies; for example, \citet{1985A&A...143..148H} reported H$_2$CO observations conducted with the 100$\,$m Effelsberg Telescope toward six sources, including G29.9$-$0.0, which is a pointing position near G29.96$-$0.02 ($\sim 2.5\arcmin$ offset). They report a [H$_2^{12}$CO]/[H$_2^{13}$CO] ratio of 45 $\pm$ 5, which is very similar to the value we obtained for G29.96$-$0.02.

The [H$_2^{12}$CO]/[H$_2^{13}$CO] value corresponds to the [$^{12}$C]/[$^{13}$C] isotope ratio if no chemical differentiation between atomic [$^{12}$C]/[$^{13}$C] and formaldehyde [H$_2$CO]/[H$_2^{13}$CO] abundances occurs\footnote{See \citet{1976A&A....51..303W} for assumptions in the determination of the [$^{12}$C]/[$^{13}$C] isotope ratio based on H$_2^{13}$CO and H$_2$CO observations.}. However, a number of studies have reported that $^{13}$C abundance can be significantly lower in formaldehyde with respect to CO, which could be due to different chemical production paths of formaldehyde \citep{1976ApJ...205L.165W, 1982ApJ...254..538K, 1984ApJ...277..581L, 2012LPI....43.1611W}. Keeping in mind the large uncertainty, the isotopic ratio we estimate in NGC$\,$7538 is similar to the values obtained for sources in the Galactic Center (e.g., \citealt{1990ApJ...357..477L}; \citealt{2014A&A...570A..65G}), even though NGC$\,$7538 is located at a galactocentric distance of 9.8$\,$kpc (derived from \citealt{2009ApJ...693..406M}). If confirmed, this result would show that enhancements in H$^{13}$CO can be found far from the Galactic Center\footnote{We note that \citet{2016A&A...593A..46F} recently derived H$_2$CO column densities assuming a smaller H$_2^{13}$CO abundance in NGC$\,$7538 S by using [$^{12}$C]/[$^{13}$C] = 73 \citep{2014A&A...570A..65G}.}. Interferometric mapping and higher signal-to-noise observations of both isotopologues and radio continuum (with matching $(u,v)$ coverage) are necessary to reduce the uncertainties of the [H$_2^{12}$CO]/[H$_2^{13}$CO] determinations for NGC$\,$7538 and the other sources in our sample.

\subsection{Methanol}

\subsubsection{12.2$\,$GHz CH$_3$OH Masers in NGC$\,$7538$\,$IRS$\,$1}

The 12.2$\,$GHz methanol masers in NGC$\,$7538$\,$IRS$\,$1 were first detected in 1985 by \citet{1987Natur.326...49B} and re-observed in 1987 by \citet{1988ApJ...326..931K}. The maser was characterized by a main velocity component at $-56.5$\kms~and weaker components at a velocity of $\sim -61.2$\kms. Long-term monitoring observations of the 12.2$\,$GHz methanol masers in NGC$\,$7538$\,$IRS$\,$1 were presented by \citet{2012IAUS..287..186P}. They reported a linear decrease in flux density of the main velocity component ($-$56.5~\kms) during a period of almost two decades (6 VLBA observing epochs between 1995 and 2005, and the earlier data from \citet{1987Natur.326...49B} and \citet{1988ApJ...326..931K}). The variability is characterized by a slope of $\sim -5.4\,$Jy$\,$year$^{-1}$. Our observations suggest that this trend has continued (at least until May 2010, Table \ref{tab:tbl-6}), in particular, the flux density decrease between our 2008 and 2010 observations is $-3 \pm 2\,$Jy$\,$year$^{-1}$ (error obtained by assuming a 5\% flux density calibration error, see section 2.1.2), and the extrapolation of the \citet{2012IAUS..287..186P} linear decrease matches our observations within 7$\%$. \citet{2012IAUS..287..186P} mentioned that the variability trend of the main 12.2$\,$GHz methanol component and possible spatial and spectral narrowing of the maser feature suggest saturated emission. We note that the velocity of the main 12.2$\,$GHz methanol velocity component is similar to the velocity of the 6$\,$cm H$_2$CO Component I (i.e., $-$56~\kms~and $-$57.8\kms, respectively), however, the masers are not spatially coincident (the main 12.2$\,$GHz methanol maser is located at the 6.7$\,$GHz methanol region A, Figure~\ref{fig:ngc7538_cont}; see also Figure~2 of \citealt{2002A&A...383..614M}).

\citet{1991ApJ...380L..75M} detected the 6.7$\,$GHz methanol line in NGC$\,$7538$\,$IRS$\,$1, which differed from the line profile of the 12.2$\,$GHz line. Specifically, a 6.7$\,$GHz methanol maser component at $\sim -58$\kms~ was not present in the 12.2$\,$GHz methanol spectra of \citet{1987Natur.326...49B} and \citet{1988ApJ...326..931K}. The absence of the $\sim -58$\kms~velocity component in the 12.2$\,$GHz methanol spectra is also seen in Figure~1 of \citet{2000A&A...362.1093M} based on VLBA observations conducted between 1997 and 1999. As shown in our spectra (Figure \ref{fig:f4}), several 12.2$\,$GHz methanol maser features are present at $\sim -58$\kms~in our 2008 and 2010 data, with flux densities of the order of 10$\,$Jy. \citet{2009ApJ...693..406M} also reported 12.2$\,$GHz methanol masers at $\sim -58$\kms~based on VLBA observations conducted in 2005 and 2006, but they reported smaller flux densities (0.1$\,$Jy). The flux density difference between our 2008 and 2010 observations and \citet{2009ApJ...693..406M} values could be caused by spatial-filtering of 12.2$\,$GHz signal by the VLBA in addition to intrinsic maser variability. In contrast to the early 12.2 and 6.7$\,$GHz methanol detections, all velocity components in our 2008 and 2010 12.2$\,$GHz methanol spectra (Figure \ref{fig:f4}) have corresponding features in more recent 6.7$\,$GHz methanol spectra, e.g., the spectra obtained with MERLIN in 2005 and EVN in 2009 \citep{2011A&A...533A..47S}. Thus, with the appearance of the $-58$\kms~features, the 12.2$\,$GHz methanol maser profile in NGC$\,$7538$\,$IRS$\,$1 has changed from showing clear differences with respect to the 6.7$\,$GHz profile, to becoming remarkably similar within two decades. According to the statistical study of \citet{2011ApJ...733...80B}, sources with both 6.7 and 12.2$\,$GHz methanol masers may be more evolved than regions with only 6.7$\,$GHz masers. Follow up observations of the 12.2$\,$GHz methanol masers in the region are required to explore whether their emergence from 1986 to $\sim$2005 is indicative of stochastic variability, which would cast doubts on the reliability of using 6.7 and 12.2$\,$GHz methanol masers as evolutionary indicators.

\subsubsection{Methanol Absorption} 

We detected two K-Band methanol transitions: 20.9706510$\,$GHz (torsionally excited, $E_{up}$ = 452$\,$K) and 20.1710890$\,$GHz (torsional ground state, $E_{up}$ = 166$\,$K) in three observing runs (epochs 1, 2, and 5; see Table~\ref{tab:tbl-6} and Figure~\ref{fig:f4}). The line parameters are similar to those reported by \citet{1986A&A...169..271M}, and the apparent variability of the line parameters seen in our data is likely caused by calibration and pointing variations in addition to the large rms (Figure~\ref{fig:f4}). We also report detection of weak absorption ($-12\,$mJy at $-$60\kms; rms = 3$\,$mJy) of the 14.782212$\,$GHz $^{13}$CH$_3$OH isotopologue toward NGC$\,$7538$\,$IRS$\,$1. We detected the weak line in two epochs (epochs 1 and 5, see Figure \ref{fig:f4}). This transition corresponds to the 12.2$\,$GHz line of the main methanol isotopologue, thus, it is very likely that the 12.2$\,$GHz methanol line also has absorption in NGC$\,$7538$\,$IRS$\,$1, but the absorption is blended with the strong 12.2$\,$GHz methanol masers and thus could not be detected.

We note that additional thermal methanol transitions have been recently observed towards NGC\,7538\,IRS1 in the millimeter range \citep{2013A&A...558A..81B, 2016A&A...592A..21F}. Interestingly, while some of the lower excited transitions were seen in absorption, several other transitions showed emission lines (see \citealt{2016A&A...592A..21F}, their Table~5). No clear trend for absorption or emission could be seen with regard to A- or E-symmetry, or regarding torsionally excited vs. ground state. But it appeared that transitions with upper-level energies $>$250$\,$K usually show emission. This tentative trend is not obeyed in our data where the 20.97$\,$GHz transition appears in absorption despite its high upper-level energy of 452$\,$K. One has to keep in mind, however, that our observations were conducted with a single dish telescope (GBT) while \citet{2016A&A...592A..21F} report Plateau de Bure Interferometre data, and that the underlying continuum is dominated in our case by free-free emission, while for the observations listed in \citet{2016A&A...592A..21F}, the continuum is a mixture of free-free emission with a steeply rising contribution from thermal dust. The small-scale brightness temperature distribution in the continuum is different between the two regimes.

\subsection{Water Masers}

Our three epoch spectra of the water masers in NGC$\,$7538$\,$IRS$\,$1 (2008, 2009, and 2010; Figure \ref{fig:f5}) exemplifies the well-known characteristic of high level of variability of water masers (e.g., \citealt{2015A&A...575A..49C}, \citealt{2007ARep...51...27L}, \citealt{1989A&AS...79...19L}). In contrast to many of the weak components that appeared/disappeared during our observations, the brightest maser component at $-58.2$\kms~was detected in all three epochs, and showed a decrease rate in the flux density of $-20 \pm 6\,$Jy$\,$year$^{-1}$. The maser was also observed by \citet{2011AJ....142..202H} with the GBT in three epochs (2010 November 24; 2010 December 8; and 2011 January 22)\footnote{The velocities of the H$_2$O maser lines listed in \citet{2011AJ....142..202H} differ from ours by $\sim$0.6\kms~because of different assumed rest frequencies; instead of the frequency used in this work [22.23508$\,$GHz, Splatalogue], \citet{2011AJ....142..202H} used the 6(1,6)-5(2,3) F=5-4 frequency of 22.2351204$\,$GHz (LOVAS database).}. The decreasing trend of the main peak of the water maser continued to at least 2011, i.e., including the \citet{2011AJ....142..202H} data with ours, the slope of the variability trend is $-33 \pm 6\,$Jy$\,$year$^{-1}$. This variability slope is steeper than the one of the 12.2$\,$GHz methanol maser ($-3 \pm 2\,$Jy$\,$year$^{-1}$, see Sect. 4.2.1). 

As in the case of the main 12.2$\,$GHz methanol maser velocity component, the main component of the 22.2$\,$GHz water masers has approximately the same peak velocity ($-58.2$\kms) as Component I of the 6$\,$cm formaldehyde maser ($-57.8$\kms), both show a negative flux density rate of change, and the masers are not spatially coincident at sub-arcsecond resolution (e.g., see \citealt{2011A&A...533A..47S} and \citealt{2010ApJ...713..423G}). However, in contrast to the formaldehyde masers, significant radial velocity drift of the water maser components is observed over the years, which makes it difficult to track the evolution of individual maser components, particularly when comparing results from different telescopes and different epochs. A thorough discussion of the water maser variability and contamination of the spectra due to emission in the GBT side lobe is presented in \citet{2011AJ....142..202H} (see also \citealt{2010ARep...54..151L} and references therein).

\section{Summary}

We present results of observations of the 6$\,$cm formaldehyde masers towards NGC$\,$7538$\,$IRS$\,$1 conducted in nine epochs (2008 - 2015) with the GBT, WRST, and VLA. The main goal of the observations was to further investigate the possibility of similar variability between the two maser velocity components as proposed by \citet{2007ApJS..170..152A}. Our data show that the similar variability behavior has continued (with a time shift of $\sim 15\,$years). The similar variability may indicate that the two maser regions have similar physical dimensions and morphologies and that the maser variability could be explained by a slow change in the line-of-sight maser amplification path, perhaps related to the kinematics of the maser clouds (e.g., rotation of non-spherical maser clouds). In addition, 6$\,$cm formaldehyde masers were observed toward five other star forming regions. Four of them (G23.01$-$0.20, G23.71$-$0.20, G25.83$-$0.18, and IRAS$\,$18566+0408) showed no flux density variability (IRAS$\,$18566+0408 was observed during the quiescent phase of the maser). In G29.96$-$0.02, we detect a possible increase in flux density, however the line is blended with strong absorption, thus, interferometric observations are needed to confirm variability of this maser.

We also report observations of the H$_2^{13}$CO isotopologue of the 6$\,$cm H$_2$CO transition toward the six 6$\,$cm H$_2$CO maser regions in our sample. No emission was detected, but weak H$_2^{13}$CO absorption lines were found in all sources except G23.71$-$0.20. Due to low sensitivity and uncertainties in filling factors and excitation temperatures, a high precision determination of the [$^{12}$C]/[$^{13}$C] isotope ratio could not be obtained. However, the NGC$\,$7538 data are consistent with a low ratio, similar to Galactic Center values. Higher signal-to-noise observations are needed to confirm this possibility.

The spectra of other masers (22.2$\,$GHz water and 12.2$\,$GHz methanol masers) were also observed toward NGC$\,$7538$\,$IRS$\,$1. We found that the strongest maser components of 22.2$\,$GHz water and 12.2$\,$GHz methanol masers steadily decreased in flux density during our observations. In the case of the 22.2$\,$GHz water masers, the flux density variability of the main velocity component ($-$58.2$\,$\kms) is consistent with GBT observations of water masers in this region observed between 2010 and 2011 by \citet{2011AJ....142..202H}. We point out that the velocity of the main water maser peak is similar to the velocity of the 6$\,$cm formaldehyde maser Component I and both showed a negative flux density rate of change. In the case of the methanol masers, \citet{2012IAUS..287..186P} reported that the flux density of the main 12.2$\,$GHz maser component (approximately at $-$56.0$\,$\kms) decreased at a rate of about 5.4$\,$Jy$\,$year$^{-1}$ between 1995 and 2005. Our two measurements of the 12.2$\,$GHz masers show that such monotonic variability continued until at least May 2010. 

Previous observations of the 12.2$\,$GHz methanol masers towards NGC$\,$7538$\,$IRS$\,$1 over the last two decades showed that some velocity components were not present at the velocities of some of the 6.7$\,$GHz maser peaks, and thus, the spectral line profiles of 12.2 and 6.7$\,$GHz masers were different. Our data now show that these missing 12.2$\,$GHz velocity components appeared and increased in flux density, and the overall spectral profiles of both transitions are now very similar. We also detected two other methanol transitions in absorption at 20.17 and 20.97$\,$GHz, and weak absorption of $^{13}$CH$_3$OH at 14.8$\,$GHz, which corresponds to the 12.2$\,$GHz transition of the main isotopologue. Therefore, it is very likely that the main isotopologue also shows absorption, but it was not detected because of spectral blending with strong 12.2$\,$GHz methanol masers.

\acknowledgments

We thank the anonymous referee for constructive comments that helped us improve the manuscript. The National Radio Astronomy Observatory is a facility of the National Science Foundation operated under cooperative agreement by Associated Universities, Inc. I.M.H. thanks the WSRT staff for flexible scheduling of the array and thanks S. R. DeSoto for assistance in data reduction. P.H. acknowledges partial support from NSF grant AST-0908901. This work has made use of the computational facilities donated by Frank Rodeffer to the Astrophysics Research Laboratory of Western Illinois University. We also acknowledge a former WIU student, Edita Ezerskyte, for participation in early stages of this project. This research made use of the NASA's Astrophysics Data System and SIMBAD.

\software{GBTIDL, AIPS \citep{1996ASPC..101...37V}, CASA \citep{2007ASPC..376..127M}, CLEO}.

\end{document}